\documentclass[aps,prd,preprint,superscriptaddress,showpacs,nofootinbib]{revtex4}

\usepackage{graphicx}
\usepackage{amsfonts,amsmath,amssymb,amsthm}
\usepackage{xcolor}

\usepackage{enumitem}
\usepackage{array}
\usepackage{multirow}
\usepackage{longtable}

\makeatletter
\newcommand\newtag[2]{#1\def\@currentlabel{#1}\label{#2}}
\makeatother

\def\be{\begin{equation}}
\def\ee{\end{equation}}
\def\ba{\begin{eqnarray}}
\def\ea{\end{eqnarray}}

\begin{document}

\title{Generalized CP Symmetries in Three-Higgs-Doublet Models}

\author{Iris Bree\footnote{ iris.bree.silva@tecnico.ulisboa.pt}}
\affiliation{CFTP, Departamento de F\'{i}sica, Instituto Superior T\'{e}cnico,
Universidade de Lisboa,\\
Avenida Rovisco Pais 1, 1049 Lisboa, Portugal
}
\author{Duarte D. Correia\footnote{ duarte.d.correia@tecnico.ulisboa.pt}}
\affiliation{CFTP, Departamento de F\'{i}sica, Instituto Superior T\'{e}cnico,
Universidade de Lisboa,\\
Avenida Rovisco Pais 1, 1049 Lisboa, Portugal
}
\author{Jo\~{a}o P. Silva \footnote{jpsilva@cftp.ist.utl.pt}}
\affiliation{CFTP, Departamento de F\'{i}sica, Instituto Superior T\'{e}cnico,
Universidade de Lisboa,\\
Avenida Rovisco Pais 1, 1049 Lisboa, Portugal
}

\date{\today}

\begin{abstract}
We study the scalar and Yukawa sectors of
three Higgs doublets models with a generalized CP symmetry.
We show that there are only four classes of scalar potentials,
merely one more than in two Higgs doublet models (2HDM).
In 2HDM with generalized CP symmetries extended to the Yukawa sector,
there are only two possible cases:
the usual CP, with 18 real Yukawa couplings;
and a minimal generalized CP model,
with 12 real Yukawa parameters.
In contrast,
with three Higgs there is a rich variety of allowed models.
We classify all possible Yukawa textures,
showing that there are 40 possibilities,
several of which have only 10 real Yukawa couplings.
\end{abstract}


\maketitle

\section{\label{sec:intro}Introduction}

The scalar sector responsible for
electroweak symmetry breaking is currently being incisively probed.
Since there is no conclusive theoretical argument constraining
the number of Higgs doublets,
ultimately it should be determined experimentally.

However,
models with more than one Higgs introduce many new parameters,
both in the scalar potential and in the Yukawa couplings.
These can be reduced with the introduction of extra
symmetries of the type
\be
\Phi_a \rightarrow S_{ab} \Phi_b,
\label{FS}
\ee
or
\be
\Phi_a \rightarrow X_{ab} \Phi_b^\ast,
\label{GCP}
\ee
where $S$ and $X$ belong to $SU(n_H)$,
$n_H$ is the number of Higgs doublets in the theory, and,
unless stated otherwise, summation of repeated indices
will always be implied.
The former are known as family symmetries.
The latter combine unitary transformations with the
usual charge-parity (CP) conjugation and were named generalized CP (GCP) transformations,
first analyzed by Lee \cite{Lee:1966ik}.
Their use in the scalar sector was first developed by
the Vienna group \cite{Ecker:1981wv, Ecker:1983hz, Neufeld:1987wa}
and their explicit use for quarks first appeared in \cite{Bernabeu:1986fc}.

It has been shown that applying the GCP symmetry
to the 2HDM scalar potential with any possible choice of $X$
leads only to three classes of scalar potentials \cite{Ivanov:2007de,Ferreira:2009wh}.
Extending that symmetry to the Yukawa sector
involves many new parameters and one might expect
a bonanza of possibilities.
As shown by Ferreira and Silva
\cite{Ferreira:2010bm},
quite the contrary happens;
besides the usual CP conserving 2HDM,
there is only one single GCP symmetric 2HDM with both
scalar and Yukawa interactions which is consistent
with nonzero quark masses.
It turns out that that model exhibits a new type of
spontaneous CP violation \cite{Ferreira:2010bm};
one where the scalar potential by itself is
CP conserving (even after spontaneous electroweak symmetry
breaking),
but where the relative phase of the vacua induces
CP-violating phases in the coupling to fermions
and in the CKM matrix.
It is thus interesting to see which of these features
are characteristic of 2HDM and which survive when there
are extra Higgs doublets.
To this end we study the scalar sector and the Yukawa sector
of three Higgs doublet models (3HDM) with a generalized
CP symmetry.

The implications of family symmetries to
the three Higgs scalar potential
were studied in \cite{Ferreira:2008zy},
for the case of symmetry groups with only one generator.
Still regarding the scalar potential,
the set of all symmetry/constrained 3HDM was mapped in
\cite{Ivanov:2012fp}\footnote{One of us (I.B.) proved later
that there are a few groups that should be added to the list;
such as $U(1)\times U(1) \rtimes \mathbb{Z}_3$ and $O(2) \times U(1)$
\cite{Tese_Iris}.},
their identification in a basis invariant fashion proposed in
\cite{Ivanov:2018ime,deMedeirosVarzielas:2019rrp},
a detailed description of their symmetry breaking patterns discussed in
\cite{Ivanov:2014doa},
and their decoupling limit properties established in \cite{Carrolo:2021euy}.
As for the impact of flavour symmetries on the combined Higgs and Yukawa sector
of the 3HDM,
it was first tackled for the case of symmetry groups with only one generator in
\cite{Ferreira:2010ir}, and later generalized into other groups in
\cite{Serodio:2013gka, Ivanov:2013bka}.

Here,
we concentrate exclusively on GCP symmetries.
In Section~\ref{sec:scalar} we employ a standard method to outline a proof
that there are only four types of GCP-symmetric 3HDM scalar potentials.
In Section~\ref{sec:new} we propose a much faster and more elegant method
to acertain symmetry constraints,
and we analyze the scalar potential, now in full detail.
The new method is crucial when we extend GCP into the
Yukawa sector in Section~\ref{sec:yukawa}.
We show explicitly all possible Yukawa textures in
Section~\ref{sec:yukawa_matrices}.
It turns out that there are many textures with fewer parameters
than present in the GCP-symmetric 2HDM.
We present our conclusions in Section~\ref{sec:conclusions}.
In
each section we include only simple examples of the type
of analysis required,
so that the reasoning leading to the conclusions stated is clear.
We relegate some details to the appendices.
In Appendix~\ref{app:scalar_matrices} we show how one reaches
the conclusion that there are only four GCP-symmetric 3HDM scalar potentials
using a standard (inefficient) analysis.
In Appendix~\ref{app:torus} we state a mathematical result
which greatly simplifies the analysis, whose detailed proof we
supply as supplemental material.

\section{\label{sec:scalar}The scalar potential}

\subsection{Notation}

Let us consider a three Higgs doublet model
(3HDM) with 3 Higgs-doublets $\Phi_i$,
of the same hypercharge $1/2$,
and with vacuum expectation values (vevs)
\begin{equation}
\langle \Phi_i \rangle
=
\left(
\begin{array}{c}
\langle \phi^+_i \rangle\\
\langle \phi^0_i \rangle
\end{array}
\right)
=
\left(
\begin{array}{c}
0\\
v_i/\sqrt{2}
\end{array}
\right)\, .
\label{vev}
\end{equation}
The index $i$ runs from $1$ to $3$,
and we use the standard definition for the electric
charge, 
whereby the upper components of the $SU(2)$ doublets ($\phi_i^+$) are
charged and the lower components ($\phi_i^0$) are neutral.

The most general 3HDM scalar potential which is renormalizable
and compatible with the gauge symmetries of the SM,
can be written as \cite{Botella:1994cs,Branco:1999fs,Davidson:2005cw}
\begin{equation}\label{eq:sclptl}
    V_H = Y_{ij}(\Phi_i^\dagger\Phi_j) +
Z_{ij,kl}(\Phi_i^\dagger\Phi_j)(\Phi_k^\dagger\Phi_l)\,\text{,}
\end{equation}
where $Y$ ($Z$) is a rank-2 (rank-4) tensor in 3-dimensions
and $Z_{ij,kl}\equiv Z_{kl,ij}$.
Hermiticity implies that
\begin{eqnarray}
Y_{ij} &=& Y_{ji}^\ast,
\nonumber\\
Z_{ij,kl} \equiv Z_{kl,ij} &=& Z_{ji,lk}^\ast\, .
\label{Hermiticity_coefficients}
\end{eqnarray}
This means that there are only 3 real (3 complex) parameters
in $Y$ and 9 real (18 complex) parameters in $Z$.

\subsection{GCP symmetries}

The scalar potential in Eq.~\eqref{eq:sclptl} is invariant
under the GCP transformation in Eq.~\eqref{GCP} if and only if
\begin{equation}\label{eq:sclinv}
    \begin{split}
    & Y_{ab}^* = X_{\alpha a}^* Y_{\alpha\beta} X_{\beta b} = ( X^\dagger\, Y\, X )_{ab}\, , \\
    & Z_{ab,cd}^* = X_{\alpha a}^* X_{\gamma c}^*
Z_{\alpha\beta,\gamma\delta} X_{\beta b} X_{\delta d}\,\text{. }
    \end{split}
\end{equation}
Solving these equations for every (independent) parameter
in $Y$ and $Z$ is a daunting task if we use a general $3\times3$
unitary matrix in Eq.~\eqref{GCP}.
Fortunately, Ecker, Grimus, and Neufeld \cite{Ecker:1987qp}
proved that there is always a basis of scalar fields,
for which the GCP transformation matrix $X$ may be brought to the form
\begin{equation}\label{eq:basis}
    X_\theta = \begin{bmatrix}
        c_\theta & s_\theta & 0 \\
        -s_\theta & c_\theta & 0 \\
        0 & 0 & 1
    \end{bmatrix} \equiv R_\theta\oplus 1\, \text{,}
\end{equation}
where the $\oplus$ symbol stands for direct sum, and $0\leq\theta\leq\pi/2$.
Notice the restricted range for $\theta$.
Henceforth,
\be
R=\begin{pmatrix}
    c & s \\
    -s & c
\end{pmatrix}\, ,
\ee
where $c=\cos$, $s=\sin$ and the Greek subindices
indicate the angle.

\section{\label{sec:new}A new strategy for GCP constraints}

\subsection{\label{subsec:simple}A simple example}

Consider a vector of complex entries $(a,b)^\top$, and a system of two equations
\be
\left(
\begin{array}{c}
a\\
b
\end{array}
\right)^*
=
R^\top_\theta\, 
\left(
\begin{array}{c}
a\\
b
\end{array}
\right)\, ,
\ee
which may be written alternatively as $(a, b)^\dagger = R_\theta^\top (a, b)^\top$.
The system has 3 distinct solutions:
\begin{eqnarray}
i) & & \textrm{if } \theta=2k\pi, \textrm{for integer } k, \textrm{then }  a,b\in\mathbb{R};
\nonumber\\
ii) & & \textrm{if } \theta=(2k+1)\pi, \textrm{for integer } k, \textrm{then } a,b\in\mathbb{I};
\nonumber\\
iii) & & \textrm{otherwise}, a=b=0.
\label{3-cases}
\end{eqnarray}
This will be the source of the subsequent analysis, provided one can
turn problems into two-dimensional blocks.

Imagine now that there is a 4-vector subject to the constraint
\be
\left(
\begin{array}{c}
a\\
b\\
c\\
d
\end{array}
\right)^*
=
R_\alpha^\top\otimes R_\beta^\top\, 
\left(
\begin{array}{c}
a\\
b\\
c\\
d
\end{array}
\right)
=
\left(
\begin{array}{cccc}
 c_\alpha c_\beta & -c_\alpha s_\beta & -c_\beta s_\alpha &
   s_\alpha s_\beta \\
 c_\alpha s_\beta & c_\alpha c_\beta & -s_\alpha s_\beta &
   -c_\beta s_\alpha \\
 c_\beta s_\alpha & -s_\alpha s_\beta & c_\alpha c_\beta &
   -c_\alpha s_\beta \\
 s_\alpha s_\beta & c_\beta s_\alpha & c_\alpha s_\beta &
   c_\alpha c_\beta \\
\end{array}
\right)\, 
\left(
\begin{array}{c}
a\\
b\\
c\\
d
\end{array}
\right)\, ,
\label{eq_4v}
\ee
where $\otimes$ denotes the Kronecker product of two matrices.
One interesting way to solve Eq.~\eqref{eq_4v} is the following.
Consider the orthogonal matrix
\begin{equation}
    C_2 = \frac{1}{\sqrt{2}}\begin{pmatrix}
        \ 1\  & \ 0\  & 0 &-1 \\
        0 & 1 & 1 & 0 \\
        1 & 0 & 0 & 1 \\
        0 & 1 & -1& 0 
    \end{pmatrix}\, .
\label{C2_definition}
\end{equation}
It is easy to show that 
\begin{equation}\label{C2}
C_2 R_\alpha^\top\otimes R_\beta^\top C_2^\top
= R_{\alpha+\beta}^\top \oplus  R_{\beta - \alpha}^\top\, .
\end{equation}
Again, the $\oplus$ symbol stands for direct sum and, here, it means
that a $4 \times 4$ block-diagonal matrix has been built,
such that the $2 \times 2$ upper-left corner has the matrix $R_{\alpha+\beta}^\top$,
the  $2 \times 2$ lower-right corner has the matrix $R_{\beta - \alpha}^\top$,
and all other entries vanish.
Left multiplying Eq.~\eqref{eq_4v} by $C_2$ and inserting $C_2^\top C_2$ in between
$R_\beta^\top$ and $C_2^\top$ on the left-hand-side (LHS) of Eq.~\eqref{C2},
one finds
\be
\left(
\begin{array}{c}
a-d\\
b+c\\
a+d\\
b-c
\end{array}
\right)^*
=
\left(
\begin{array}{cc}
R_{\alpha+\beta}^\top & O\\
O & R_{\beta - \alpha}^\top
\end{array}
\right)\, 
\left(
\begin{array}{c}
a-d\\
b+c\\
a+d\\
b-c
\end{array}
\right)\, ,
\label{eq_4v_new}
\ee
effectively decoupling the problem into much simpler two two-dimensional problems.
But now one can use Eq.~\eqref{3-cases} for immediate conclusions.
This is generalized to higher dimensions in Appendix~\ref{app:torus}.

It will prove useful to use the vec($\cdot$) operator, which vectorizes by rows.
It can be shown that
\begin{eqnarray}
\text{vec}(C_{ck} \bar{Y}_{kl}D_{ld})
&=& C\otimes D^\top \text{vec}(\bar{Y}_{kl})
\label{ordem1}
\\
\text{vec}(A_{ai}C_{ck}\bar{K}_{i,kl}D_{ld})
&=& A\otimes C\otimes D^\top \text{vec}(\bar{K}_{i,kl})
\label{ordem2}
\\
\text{vec}(A_{ai}C_{ck}\bar{Z}_{ij,kl}B_{jb}D_{ld})
&=&
A\otimes B^\top \otimes C \otimes D^\top \text{vec}(\bar{Z}_{ij,kl})\,\text{,}
\label{ordem3}
\end{eqnarray}
where $A$, $B$, $C$ and $D$ are appropriately sized matrices,
vec($\cdot$) vectorizes matrices by row, $\otimes$ denotes
the Kronecker product,
whereas $\bar Y$, $\bar K$ and
$\bar Z$ are tensors of rank 2, 3 and 4, respectively.

\subsection{GCP constraints on $Y$ }

Here, we apply the technique of the previous section in order to solve
Eqs.~\eqref{eq:sclinv} and derive
the GCP constraints on $Y$. We include the study of the GCP symmetry
conditions on the scalar potential using a different strategy,
which mimics \cite{Ferreira:2010bm}, in Appendix~\ref{app:scalar_matrices}.
Employing the strategy in Appendix~\ref{app:scalar_matrices} to the Yukawa sector,
however, becomes greatly error-prone.

Due to the simple form of Eq.~\eqref{eq:basis},
the tensor is divided into 4 regions:
$mn$, $m3$, $3n$, and $33$ $(m,n=1,2)$.
But, thanks to Hermiticity, we need not solve the system for $3n$.
For $33$, we have that $Y_{33}^* = Y_{33}\Leftrightarrow Y_{33} \in \mathbb{R} $,
which we already knew.

Henceforth, we will be using the notation
\begin{equation}\label{eq:notation}
\begin{split}
    Y_{\{m3\}}&=(Y_{13},Y_{23})^\top\, ,\\
    Y_{\{3n\}}&=(Y_{31},Y_{32})^\top\, ,\\
    Y_{\{mn\}}&=(Y_{11},Y_{12},Y_{21},Y_{22})^\top\, .
\end{split}
\end{equation}
For $m3$ and $3n$, we have 
\begin{equation}\label{eq:r}
    Y_{\{m3\}}^* = R_\theta^\top Y_{\{m3\}}\, ,
    \;\;\;\;
    Y_{\{3n\}}^* = R_\theta^\top Y_{\{3n\}}\, .
\end{equation}
As seen in Eqs.~\eqref{3-cases} and given the restricted range for $\theta$, we find that

\bgroup
\def\arraystretch{1.5}
\begin{tabular}{rp{15mm}l}
i. & $\theta=0$ & $(Y)_{13}=(Y)_{31}$, $(Y)_{23}=(Y)_{32}$  $\in\mathbb{R}\, ,\,$ \\
ii. & $\theta\neq0$ & $(Y)_{13}=(Y)_{31}=(Y)_{23}=(Y)_{32}=0\,.\,$   \\
\end{tabular}
\egroup

\vspace{2mm}

\noindent
For $mn$, we use Eq.~\eqref{ordem1},
in order to rewrite Eq.~\eqref{eq:sclinv} as
\begin{equation}\label{eq:rr}
    Y_{\{mn\}}^* = R_\theta^\top\otimes R_\theta^\top Y_{\{mn\}}\, .
\end{equation}
To solve this system we use Eq.~\eqref{C2}.
The $mn$ sector then simplifies to
\begin{equation}\label{eq:11}
    \begin{split}
         \begin{pmatrix}
            Y_{11} - Y_{22} \\
            2\,\text{Re}(Y_{12})
        \end{pmatrix}^*&=R_{2\theta}^\top
        \begin{pmatrix}
            Y_{11} - Y_{22} \\
            2\,\text{Re}(Y_{12})
        \end{pmatrix}\,\text{,} \\
        \text{Im}(Y_{12}) &= 0\,\text{,}
    \end{split}
\end{equation}
and solving it results in the following two options:

\bgroup
\def\arraystretch{1.5}
\begin{tabular}{rp{15mm}l}
i. & $\theta=0$ & $(Y)_{11}$, $(Y)_{22}$, $(Y)_{12}=(Y)_{21}$  $\in\mathbb{R}\, ,\,$ \\
ii. & $\theta\neq0$ & $(Y)_{11}=(Y)_{22}$, $(Y)_{12}=(Y)_{21}=0\,.\,$   \\
\end{tabular}
\egroup

\vspace{2mm}

\noindent
In conclusion:
for $\theta=0$, all of $Y$'s entries are real, otherwise $Y=\text{diag}(\mu_1,\mu_1,\mu_3)$.

\subsection{Summary of GCP constraints on $Z$}

The full simplifying power of the new method comes to light
when we deal with the quartic $Z$ couplings.\footnote{We included the messier alternative
in Appendix~\ref{app:scalar_matrices} for comparison.}
As before, our choice of basis decouples the third entries
from the rest, but now we have a total of 16 separate regions,
those being $ijkl$, $ijk3$, $ij3l$, $i3kl$, $3jkl$, $ij33$, $i3k3$,
$i33l$, $3jk3$, $3j3l$, $33kl$, $i333$, $3j33$, $33k3$,
$333l$ and $3333$ (here $i,j,k,l = 1,2$).
And thanks once again to Hermiticity, we need not solve all of them.
We just need to solve 7: $ijkl$, $ij33$, $i3kl$, $i3k3$, $i33l$,
$i333$ and $3333$, since these contain all independent entries for $Z$. 

The case for $3333$ just tells us that $Z_{3333}$ is a real entry,
which we already knew.
For $i333$, we will have the same as Eqs.~\eqref{eq:r},
and so these entries are always real,
and equal to 0 if $\theta\neq 0$.
The sectors $ij33$, $i33l$ and $i3k3$
will be the same as Eq.~\eqref{eq:rr}. 
Although the first two cases are identical to Eq.~\eqref{eq:11},
the latter is a bit more interesting,
as we discover a new region of interest when $\theta=\pi/2$.

Finally,
we have $Z_{\{i3kl\}}^*= {R_\theta^\top}^{\otimes 3}Z_{\{i3kl\}}$
and $Z_{\{ijkl\}}^*= {R_\theta^\top}^{\otimes 4}Z_{\{ijkl\}}$,
with a definition analogous to Eq.~\eqref{eq:notation}.
As before, there are orthogonal matrices $C_3$ and $C_4$ such that
\begin{equation}\label{C3}
    C_3{R_\theta^\top}^{\otimes 3} C_3^\top
= R_{3\theta}^\top\oplus R_\theta^\top\oplus
R_\theta^\top \oplus R_{\theta}\, ,
\end{equation}
and
\begin{equation}
    C_4 {R_\theta^\top}^{\otimes 4} C_4^\top 
    = R_{4\theta}^\top\oplus R_{2\theta}^\top\oplus
 R_{2\theta}^\top \oplus I_2 \oplus R_{2\theta}^\top\oplus
I_2\oplus I_2 \oplus R_{2\theta}\, .
\end{equation}
This technique is quite useful since it turns a
$R_\theta^{\otimes n}$ matrix, a $2^n\times2^n$ system,
into $2^{n-1}$ $2\times2$ systems,
which are trivial to solve.
The $i3kl$ system provides us with the final region of interest,
that being when $\theta=\pi/3$.

At last, we have our four regions of interest:
$\theta=0$, $\theta=\pi/2$, $\theta=\pi/3$ and
$\theta\in(0,\pi/2)\setminus\{\pi/3\}$,
which we'll denote as
CPa, CPb, CPc and
CPd, respectively\footnote{In
the notation of the 2HDM case in \cite{Ferreira:2010bm},
these would be written as, respectively, CP1, CP2,
with both c and d corresponding there to
CP3.
}.
Let us compare this result with what one has in the 2HDM.
In the scalar potential of the 2HDM there are 2 real and
1 complex parameters in the quadratic couplings and 4 real and
3 complex parameters in the quartic couplings.
Imposing GCP we find only three classes of 2HDM potentials,
corresponding to $\theta=0$, $\theta=\pi/2$, and $\theta\in(0,\pi/2)$.
In the 3HDM there are 3 real and 3
complex parameters in the quadratic couplings and 9 real and 18 complex
parameters in the quartic couplings.
Despite the enormous increase in the number of parameters,
there is only one more class of GCP-symmetric potentials:
corresponding to the singling out of the
$\theta=\pi/3$ case.

\subsection{Proof of GCP constraints on $Z$}

Using Eq.~\eqref{ordem3}, we can write
\begin{equation}\label{eq:A8}
        Z_{3333}^*
=Z_{3333}\, ,
\end{equation}
%
%
\begin{equation}\label{eq:A9}
        Z_{\{i333\}}^*= R_\theta^\top Z_{\{i333\}}\, ,
\ \ \ \  Z_{\{i333\}}=(Z_{1333}, Z_{2333})^\top\, ,
\end{equation}
%
%
\begin{equation}\label{eq:A10}
    \begin{split}
        Z_{\{ij33\}}^*= R_\theta^{\top\otimes 2} Z_{\{ij33\}}\, ,
&\ \ \ \  Z_{\{ij33\}}=(Z_{1133}, Z_{1233}, Z_{2133}, Z_{2233})^\top\, , \\     
        Z_{\{i33l\}}^*= R_\theta^{\top\otimes 2} Z_{\{i33l\}}\, ,
&\ \ \ \  Z_{\{i33l\}}=(Z_{1331}, Z_{1332}, Z_{2331}, Z_{2332})^\top\, , \\
        Z_{\{i3k3\}}^*= R_\theta^{\top\otimes 2} Z_{\{i3k3\}}\, ,
&\ \ \ \  Z_{\{i3k3\}}=(Z_{1313}, Z_{1323}, Z_{2313}, Z_{2323})^\top\, ,
    \end{split}
\end{equation}
%
%
\begin{equation}\label{eq:A11}
    \begin{split}
        Z_{\{i3kl\}}^*= R_\theta^{\top\otimes 3}
Z_{\{i3kl\}}\;\text{, } Z_{\{i3kl\}}=(
& Z_{1311}, Z_{1312}, Z_{1321}, Z_{1322}, 
Z_{2311}, Z_{2312}, Z_{2321}, Z_{2322})^\top\, ,
    \end{split}
\end{equation}
%
%
\begin{equation}\label{eq:A12}
    \begin{split}
        Z_{\{ijkl\}}^*= R_\theta^{\top\otimes 4} 
Z_{\{ijkl\}}\;\text{, } Z_{\{ijkl\}}=(
& Z_{1111}, Z_{1112}, Z_{1121}, Z_{1122}, 
Z_{1211}, Z_{1212}, Z_{1221}, Z_{1222}, \\
        & Z_{2111}, Z_{2112}, Z_{2121}, Z_{2122}, 
Z_{2211}, Z_{2212}, Z_{2221}, Z_{2222})^\top\, .
    \end{split}
\end{equation}
Eq.~\eqref{eq:A8} tell us that $Z_{3333}$ is real which we already knew.
In Eq.~\eqref{eq:A9}, due to the restricted range for
$\theta$, we have

\bgroup
\def\arraystretch{1.5}
\begin{tabular}{rp{15mm}l}
i. & $\theta=0$ & $Z_{1333}$, $Z_{2333}$  $\in\mathbb{R}\, ,\,$ \\
ii. & $\theta\neq0$ & $Z_{1333}=Z_{2333}=0\,.\,$   \\
\end{tabular}
\egroup

\vspace{2mm}

\noindent
For Eqs.~\eqref{eq:A10}, we shall use the result in Eq.~\eqref{result}
for $n=2$,
along with the following relations:
\; $Z_{1233}=Z_{2133}^*$; 
$Z_{1332}=Z_{2331}^*$;\;
$Z_{1323}=Z_{2313}$;
\; $Z_{1133}, Z_{2233}, Z_{1331},Z_{2332} \in \mathbb{R}$.
We find
\begin{equation}\label{eq:A13}
    \begin{split}
        \begin{pmatrix}
            Z_{1133} - Z_{2233} \\
            2\,\text{Re}(Z_{1233})
        \end{pmatrix}^*   &= R_{2\theta}^\top \begin{pmatrix}
                                                  Z_{1133} - Z_{2233} \\
                                                  2\,\text{Re}(Z_{1233})
                                              \end{pmatrix}\, ,\\
        \begin{pmatrix}
            Z_{1331} - Z_{2332} \\
            2\,\text{Re}(Z_{1332})
        \end{pmatrix}^*   &= R_{2\theta}^\top \begin{pmatrix}
                                                  Z_{1331} - Z_{2332} \\
                                                  2\,\text{Re}(Z_{1332})
                                              \end{pmatrix}\, ,\\
        \text{Im}(Z_{1233}) &= \text{Im}(Z_{1332})=0\, ,
    \end{split}
\end{equation}
\begin{equation}\label{eq:A14}
    \begin{split}
        \begin{pmatrix}
            Z_{1313} - Z_{2323} \\
            2\,Z_{1323}
        \end{pmatrix}^*   &= R_{2\theta}^\top \begin{pmatrix}
                                                  Z_{1313} - Z_{2323} \\
                                                  2\,Z_{1323}
                                              \end{pmatrix}\, ,\\*[1mm]
         Z_{1313} &+ Z_{2323} \in \mathbb{R}\, .
    \end{split}
\end{equation}
For Eqs.~\eqref{eq:A13} we find 

\bgroup
\def\arraystretch{1.5}
\begin{tabular}{rp{15mm}l}
i. & $\theta=0$ & $Z_{\{ij33\}}$, $Z_{\{i33l\}}$  $\in\mathbb{R}\, ,\,$ \\ 
ii. & $\theta\neq0$ & $Z_{1133} = Z_{2233}$, $Z_{1331} = Z_{2332}$,
$Z_{1233} = Z_{2133} = Z_{1332}=Z_{2331}= 0\,,\,$   \\
\end{tabular}
\egroup

\vspace{2mm}

\noindent
and, similarly, for Eqs.~\eqref{eq:A14}, 

\bgroup
\def\arraystretch{1.5}
\begin{tabular}{rp{25mm}l}
i. & $\theta=0$ & $Z_{\{i3k3\}} \in\mathbb{R}\, ,\,$ \\ 
ii. & $\theta=\pi/2$ & $Z_{1323}=Z_{2313} \in\mathbb{I}$, $Z_{1313}=Z_{2323}^*$\, ,\, \\
iii. & $\theta\neq\{0,\pi/2\}$ & $Z_{1313} = Z_{2323}\in\mathbb{R}$, $Z_{1323} = Z_{2313}=0\,.\,$   \\
\end{tabular}
\egroup

\vspace{2mm}

\noindent
For Eqs.~\eqref{eq:A11}, we shall use the result in Eq.~\eqref{result}
for $n=3$. We find
\begin{equation}\label{eq:A15}
    \begin{split}
        \begin{pmatrix}
            (Z_{1311} - Z_{1322}) - (Z_{2312} + Z_{2321}) \\
            (Z_{1312} + Z_{1321}) + (Z_{2311} - Z_{2322})
        \end{pmatrix}^*   &= R_{3\theta}^\top \begin{pmatrix}
                                                 (Z_{1311} - Z_{1322}) - (Z_{2312} + Z_{2321}) \\
                                                 (Z_{1312} + Z_{1321}) + (Z_{2311} - Z_{2322})
                                              \end{pmatrix}\, ,\\
        \begin{pmatrix}
            (Z_{1311} - Z_{1322}) + (Z_{2312} + Z_{2321}) \\
            (Z_{1312} + Z_{1321}) - (Z_{2311} - Z_{2322})
        \end{pmatrix}^*   &= R_{\theta}^\top \begin{pmatrix}
                                                  (Z_{1311} - Z_{1322}) + (Z_{2312} + Z_{2321}) \\
                                                  (Z_{1312} + Z_{1321}) - (Z_{2311} - Z_{2322})
                                              \end{pmatrix}\, ,\\
        \begin{pmatrix}
            (Z_{1311} + Z_{1322}) - (Z_{2312} - Z_{2321}) \\
            (Z_{1312} - Z_{1321}) + (Z_{2311} + Z_{2322})
        \end{pmatrix}^*   &= R_{\theta}^\top \begin{pmatrix}
                                                  (Z_{1311} + Z_{1322}) - (Z_{2312} - Z_{2321}) \\
                                                  (Z_{1312} - Z_{1321}) + (Z_{2311} + Z_{2322})
                                              \end{pmatrix}\, ,\\
       \begin{pmatrix}
            (Z_{1311} + Z_{1322}) + (Z_{2312} - Z_{2321}) \\
            (Z_{1312} - Z_{1321}) - (Z_{2311} + Z_{2322})
        \end{pmatrix}^*   &= R_{\theta} \begin{pmatrix}
                                                  (Z_{1311} + Z_{1322}) + (Z_{2312} - Z_{2321}) \\
                                                  (Z_{1312} - Z_{1321}) - (Z_{2311} + Z_{2322})
                                              \end{pmatrix}\, .\\
    \end{split}
\end{equation}
The solution to these equations is

\bgroup
\def\arraystretch{1.5}
\begin{tabular}{rp{25mm}p{27mm}p{70mm}}
i. & $\theta=0$ & $Z_{\{i3kl\}} \in\mathbb{R}\, ,\,$ & \\ 
ii. & $\theta=\pi/3$ & $Z_{\{i3kl\}}= \in\mathbb{I}$,  & $Z_{1311} = -Z_{1322} = -Z_{2312} = -Z_{2321}$\,,
\newline $Z_{1312} = Z_{1321} = Z_{2311} =- Z_{2322}$\,, \\
iii. & $\theta\neq\{0,\pi/3\}$ & $Z_{\{i3kl\}} = 0\,.\,$ &  \\
\end{tabular}
\egroup

\vspace{2mm}

\noindent
For Eqs.~\eqref{eq:A12}, we shall use the result in Eq.~\eqref{result} for $n=4$
along with the following relations:
$Z_{1122}=Z_{2211}$;\; $ Z_{1221}=Z_{2112}$;\;
$Z_{1112}=Z_{1211}=Z_{1121}^*=Z_{2111}^*$;\;
$Z_{1222}=Z_{2212}=Z_{2122}^*=Z_{2221}^*$;\;
$Z_{1212}=Z_{21221}^*$;
\; $Z_{1111}, Z_{2222}, Z_{1122}, Z_{2211}, Z_{1221}, Z_{2112}\in \mathbb{R}$.
We find
\begin{eqnarray}\label{eq:A16a}
&&
\begin{pmatrix}
(Z_{1111} + Z_{2222}) - 2\,(Z_{1122} + Z_{1221} + \text{Re}(Z_{1212})) \\
4\, (\text{Re}(Z_{1112}) - \text{Re}(Z_{1222}))
\end{pmatrix}^*  
\nonumber\\
&&
\hspace{15ex}
= R_{4\theta}^\top
\begin{pmatrix}
(Z_{1111} + Z_{2222}) - 2\,(Z_{1122} + Z_{1221} + \text{Re}(Z_{1212})) \\
4\, (\text{Re}(Z_{1112}) - \text{Re}(Z_{1222}))
\end{pmatrix}\, ,
\end{eqnarray}
\begin{equation}\label{eq:A16b}
\begin{split}
      \begin{pmatrix}
            (Z_{1111} - Z_{2222}) - 2\,i\,\text{Im}(Z_{1212}) \\
            2\, (Z_{1112} + Z_{1222}^*)
      \end{pmatrix}^*   &= R_{2\theta}^\top \begin{pmatrix}
                                                 (Z_{1111} - Z_{2222}) - 2\,i\,\text{Im}(Z_{1212}) \\
                                                 2\, (Z_{1112} + Z_{1222}^*)
                                             \end{pmatrix}\, ,\\
      \begin{pmatrix}
            (Z_{1111} - Z_{2222}) + 2\,i\,\text{Im}(Z_{1212}) \\
            2\, (Z_{1112}^* + Z_{1222})
      \end{pmatrix}^*   &= R_{2\theta}^\top \begin{pmatrix}
                                                 (Z_{1111} - Z_{2222}) + 2\,i\,\text{Im}(Z_{1212}) \\
                                                 2\, (Z_{1112}^* + Z_{1222})
                                             \end{pmatrix}\, ,\\
      \text{Im}(Z_{1112}) &= -\text{Im}(Z_{1222})\, .
      \end{split}
\end{equation}
Finally, the conditions that solve these equations are

\bgroup
\def\arraystretch{1.5}
\begin{tabular}{rp{25mm}p{27mm}p{70mm}}
i. & $\theta=0$ & $Z_{\{ijkl\}} \in\mathbb{R}\, ,\,$ & \\ 
ii. & $\theta=\pi/2$ & $Z_{\{ijkl\}} \in\mathbb{C}$,  & $Z_{1112}=Z_{1211}=Z_{1121}^*=Z_{2111}^*
= -Z_{1222} = - Z_{2212}= -Z_{2122}^*
= -Z_{2221}^*$\,, \newline $Z_{1111} = Z_{2222}$\,, \\
iii. & $\theta\neq\{0,\pi/2\}$ & $Z_{\{ijkl\}}\in\mathbb{C}\,.\,$ & 
$Z_{1112}=Z_{1211}=Z_{1121}=Z_{2111} = Z_{1222} = 
Z_{2212}= Z_{2122} = Z_{2221} = 0$\,,\newline $Z_{1212} =
Z_{1111}-Z_{1122}-Z_{1221}$\,, \newline $Z_{1111} = Z_{2222}$\,. \\
\end{tabular}
\egroup

\vspace{2mm}

This concludes the analysis for the scalar sector.
If one were to try to do these calculations in models with more doublets,
the trick would remain much the same; but, then, one would have
to decouple the entries with 1 and 2 from the ones with 3 and 4, 5
and 6, etc... Since it always involves a rank-2 tensor and a
rank-4 tensor, there is no need in NHDM to go further than $C_4$.

\subsection{The four GCP-constrained 3HDM potentials}

We now write the scalar potential for the several classes of models in the basis
of Eq.~\eqref{eq:basis}.
The CPa $(\theta=0)$ scalar potential,
corresponding to the usual CP, is of the general form of
Eq.~\eqref{eq:sclptl}, except all the coefficients are real.

The CPb ($\theta=\pi/2$) scalar potential has the form
\ba
V_H
&=&
\mu_1 \left[ (\Phi_1^\dagger \Phi_1) + (\Phi_2^\dagger \Phi_2) \right]
+ 
\mu_3\, (\Phi_3^\dagger \Phi_3)
\nonumber\\
&+&
r_1 \left[ (\Phi_1^\dagger \Phi_1)^2 + (\Phi_2^\dagger \Phi_2)^2 \right]
+
r_3\, (\Phi_3^\dagger \Phi_3)^2
\nonumber\\
&+&
2\, r_4\, (\Phi_1^\dagger \Phi_1) (\Phi_2^\dagger \Phi_2)
+
2\, r_5 \left[ (\Phi_1^\dagger \Phi_1) + (\Phi_2^\dagger \Phi_2) \right] (\Phi_3^\dagger \Phi_3)
\nonumber\\
&+&
2\, r_7\, |\Phi_1^\dagger \Phi_2|^2
+
2\, r_8 \left[  |\Phi_1^\dagger \Phi_3|^2 + |\Phi_2^\dagger \Phi_3|^2 \right] 
\nonumber\\
&+&
2\, c_1 \left[ (\Phi_1^\dagger \Phi_1) - (\Phi_2^\dagger \Phi_2) \right] (\Phi_1^\dagger \Phi_2)
+
c_3\, (\Phi_1^\dagger \Phi_2)^2
+ 
\textrm{H.c.}
\nonumber\\
&+&
c_5 \left[ (\Phi_1^\dagger \Phi_3)^2 + (\Phi_3^\dagger \Phi_2)^2 \right] 
+
2\, i\, y_{11}\, (\Phi_1^\dagger \Phi_3) (\Phi_2^\dagger \Phi_3)
+
\textrm{H.c.},
\label{VH_CP2}
\ea
where H.c.~stands for Hermitian conjugation.
We follow the notation of Ref.~\cite{Ferreira:2008zy}:
the coefficients $c_k = x_k + i y_k$ are complex,
while $r_k$, $x_k$, and $y_k$ are real.

The CPc ($\theta=\pi/3$) scalar potential has the form
\ba
V_H
&=&
\mu_1 \left[ (\Phi_1^\dagger \Phi_1) + (\Phi_2^\dagger \Phi_2) \right]
+ 
\mu_3\, (\Phi_3^\dagger \Phi_3)
\nonumber\\
&+&
r_1 \left[ (\Phi_1^\dagger \Phi_1)^2 + (\Phi_2^\dagger \Phi_2)^2 \right]
+
r_3\, (\Phi_3^\dagger \Phi_3)^2
\nonumber\\
&+&
2\, r_4\, (\Phi_1^\dagger \Phi_1) (\Phi_2^\dagger \Phi_2)
+
2\, r_5 \left[ (\Phi_1^\dagger \Phi_1) + (\Phi_2^\dagger \Phi_2) \right] (\Phi_3^\dagger \Phi_3)
\nonumber\\
&+&
2\, r_7\, |\Phi_1^\dagger \Phi_2|^2
+
2\, r_8 \left[  |\Phi_1^\dagger \Phi_3|^2 + |\Phi_2^\dagger \Phi_3|^2 \right] 
\nonumber\\
&+&
r_{147} \left[ (\Phi_1^\dagger \Phi_2)^2 + (\Phi_2^\dagger \Phi_1)^2 \right]
+
x_5 \left[ (\Phi_1^\dagger \Phi_3)^2 + (\Phi_2^\dagger \Phi_3)^2  + \textrm{H.c.} \right]
\nonumber\\
&+&
2\, i\, y_2
\left[
\left[(\Phi_1^\dagger \Phi_2) + (\Phi_2^\dagger \Phi_1) \right] (\Phi_3^\dagger \Phi_2)
+ (\Phi_1^\dagger \Phi_1) (\Phi_1^\dagger \Phi_3)
+ (\Phi_2^\dagger \Phi_2) (\Phi_3^\dagger \Phi_1)
\right]
+ \textrm{H.c.},
\nonumber\\
&+&
2\, i\, y_4
\left[
\left[(\Phi_1^\dagger \Phi_2) + (\Phi_2^\dagger \Phi_1) \right] (\Phi_1^\dagger \Phi_3)
+ (\Phi_1^\dagger \Phi_1) (\Phi_2^\dagger \Phi_3)
+ (\Phi_2^\dagger \Phi_2) (\Phi_3^\dagger \Phi_2)
\right]
+ \textrm{H.c.},
\label{VH_CP3}
\ea
where $r_{147}= r_1 - r_4 - r_7$.

Finally,
the CPd potential ($\theta \neq 0, \pi/2, \pi/3$)
has the form
\ba
V_H
&=&
\mu_1 \left[ (\Phi_1^\dagger \Phi_1) + (\Phi_2^\dagger \Phi_2) \right]
+ 
\mu_3\, (\Phi_3^\dagger \Phi_3)
\nonumber\\
&+&
r_1 \left[ (\Phi_1^\dagger \Phi_1)^2 + (\Phi_2^\dagger \Phi_2)^2 \right]
+
r_3\, (\Phi_3^\dagger \Phi_3)^2
\nonumber\\
&+&
2\, r_4\, (\Phi_1^\dagger \Phi_1) (\Phi_2^\dagger \Phi_2)
+
2\, r_5 \left[ (\Phi_1^\dagger \Phi_1) + (\Phi_2^\dagger \Phi_2) \right] (\Phi_3^\dagger \Phi_3)
\nonumber\\
&+&
2\, r_7\, |\Phi_1^\dagger \Phi_2|^2
+
2\, r_8 \left[  |\Phi_1^\dagger \Phi_3|^2 + |\Phi_2^\dagger \Phi_3|^2 \right] 
\nonumber\\
&+&
r_{147} \left[ (\Phi_1^\dagger \Phi_2)^2 + (\Phi_2^\dagger \Phi_1)^2 \right]
+
x_5 \left[ (\Phi_1^\dagger \Phi_3)^2 + (\Phi_2^\dagger \Phi_3)^2  + \textrm{H.c.} \right].
\label{VH_CP4}
\ea

We have built a Mathematica program implementing the basis-invariant techniques of
Refs.~\cite{Ivanov:2018ime,deMedeirosVarzielas:2019rrp},
and checked that our CPa, CPb, CPc, and CPd potentials obey basis
invariant conditions, allowing them to be identified, respectively,
as\footnote{We are very grateful to
Igor Ivanov for several discussions on this issue,
especially pertaining to the identification of $S_3 \times GCP_{\theta=\pi}$.}
CP2,
CP4,
$S_3 \times GCP_{\theta=\pi}$,
and $O(2) \times  CP$.

Notice that, by construction,
all potentials are explicitly CP conserving.
One may wonder whether they may lead to spontaneous CP violation.
We do not address here the issue of the possible global minima when the symmetry is exact.
Aspects of spontaneous symmetry breaking in 3HDM with an exact symmetry
have been discussed in \cite{Ivanov:2014doa}.
When accessing a model's parametric viability,
we take the view that one may wish to add soft-symmetry breaking terms to the potential,
thus allowing for general $v_i$. This allows us to map all possibilities, even in that
more general context.

\section{\label{sec:yukawa}The Yukawa couplings}

\subsection{Yukawa Lagrangian and mass basis}

The scalar-quark Yukawa interactions of the 3HDM may be written as
\be
- {\cal L}_Y =
\bar{q}_L
(\Gamma_1 \Phi_1 + \Gamma_2 \Phi_2 + \Gamma_3 \Phi_3)
n_R
+
\bar{q}_L
(\Delta_1 \tilde{\Phi}_1 + \Delta_2 \tilde{\Phi}_2 + \Delta_3 \tilde{\Phi}_3)
p_R
+ \textrm{H.c.},
\label{eq:yuk}
\ee
where $q_L = (p_L, n_L)^\top$ ($n_R$ and $p_R$) is a vector in the 3-dimensional
generation space of left-handed doublets
(right-handed charge $-1/3$ and $+2/3$) quarks,
and $\Tilde{\Phi}_k = i \sigma_2 {\Phi}^*_k$, with $\sigma_2$ the second Pauli matrix.  
$\Gamma_k$ and $\Delta_k$ ($k=1,2,3$) are completely
general $3 \times 3$ complex matrices.

After spontaneous symmetry breaking,
the Lagrangian's quadratic terms in the quark fields include
\begin{equation}
        -\mathcal{L}_Y \supseteq \frac{v}{\sqrt{2}}(\overline{n}_L \Gamma n_R + \overline{p}_L \Delta p_R) + \text{H.c., }
\end{equation}
where $v=\sqrt{v_i {v_i}^*}= (\sqrt{2} G_F)^{-1/2}$, $\Gamma = \frac{v_i}{v}\Gamma_i$ and $\Delta = \frac{{v_i}^*}{v}\Delta_i$. We can define 

\begin{equation}
    M_d \equiv \frac{v}{\sqrt{2}}\Gamma\, , \ \ \ \ \ 
M_u \equiv \frac{v}{\sqrt{2}}\Delta\text{,}
\end{equation}
as the mass matrices for the down-type and up-type quarks, respectively. 
Since the fields we are working with are not mass eigenstates, these matrices will not be diagonal. 
But we can always perform a unitary change of basis
\begin{equation}
    \begin{split}
        \overline{n}_L = \overline{d}_L U_{d_L}^\dagger\, , \ \ \ \ \ 
\overline{p}_L = \overline{u}_L U_{u_L}^\dagger\,\text{, }\\
        n_R = U_{d_R} d_R\, , \ \ \ \ \ 
p_R = U_{u_R} u_R\,\text{, }
    \end{split}
\end{equation}
which leaves the kinetic terms unchanged in the Lagrangian,
to bi-diagonalize the mass matrices. The mass matrices then become
\begin{equation}
    \begin{split}
        U_{d_L}^\dagger M_d U_{d_R} &=D_d \equiv \text{diag}(m_d, m_s, m_b)\,\text{, }\\
        U_{u_L}^\dagger M_u U_{u_R} &=D_u \equiv \text{diag}(m_u, m_c, m_t)\,\text{.}
    \end{split}
\end{equation}
Writing the interaction terms of the physical fields with the $W^+$ boson, 
\begin{equation}\label{eq:22}
    \mathcal{L}_{W} \supset \frac{i g}{\sqrt{2}}\, 
\overline{u}_L (U_{u_L}^\dagger U_{d_L}) \gamma^\mu d_L\, W_\mu^+ \, ,
\end{equation}
one notices the emergence of the $3\times3$ unitary matrix
Cabibbo-Kobayashi-Maskawa (CKM) matrix \cite{Cabibbo:1963yz,Kobayashi:1973fv},
$V\equiv U_{u_L}^\dagger U_{d_L}$.
It reflects the fact that the interaction basis is distinct from the physical mass basis.
And it describes the quark mixing, responsible also for all
CP-violating phenomena in the SM.

We can also define
\begin{equation}\label{eq:HH}
    \begin{split}
        H_d \equiv  M_d  M_d^\dagger
=  \frac{v^2}{2} \Gamma \Gamma^\dagger
=U_{d_L} D_d^2 U_{d_L}^\dagger \, ,\\
        H_u \equiv  M_u  M_u^\dagger
=  \frac{v^2}{2} \Delta \Delta^\dagger
=  U_{u_L} D_u^2 U_{u_L}^\dagger\, ,
    \end{split}
\end{equation}
showing that the left-handed transformations are the
matrices that diagonalize $H_d$ and $H_u$.

\subsection{GCP symmetries for the Yukawa Lagrangian}

For the quark fields,
the GCP transformations take the form
\ba
q_L &\rightarrow X_\alpha  \gamma^0 C q^\ast_L,\nonumber \\
n_R &\rightarrow X_\beta  \gamma^0 C n^\ast_R,\nonumber \\
p_R &\rightarrow X_\gamma  \gamma^0 C p^\ast_R,
\label{GCP_fermion}
\ea
where $\gamma^0$ ($C$) is the Dirac (charge-conjugation) matrix,
and $X_\alpha$, $X_\beta$ , and $X_\gamma$ belong to $SU(3)$.
With a suitable basis choice,
these can be changed into the simplified form \cite{Ecker:1987qp}
\be
X_\alpha =
\left[
\begin{array}{ccc}
c_\alpha & s_\alpha & 0\\
- s_\alpha &  c_\alpha & 0\\
0 & 0 & 1
\end{array}
\right],
\label{eq:basis_2}
\ee
where $0 \leq \alpha \leq \pi/2$,
with similar expressions for $X_\beta$ and $X_\gamma$.

Under the transformations in Eqs.~\eqref{GCP_fermion}, the Yukawa Lagrangian becomes
\begin{equation}
-\mathcal{L}_Y
\xrightarrow{} 
        \overline{n}_R\left[ X_\alpha^\dagger (\Gamma_a )
(X_\theta)_{ab}\Phi_b^* X_\beta \right]q_L \\
+ \overline{p}_R\left[ X_\alpha^\dagger(\Delta_a)
(X_\theta)_{ab}^*\Tilde{\Phi}_b^* X_\gamma \right]q_L
+ \,\text{H.c.} \, .
\end{equation}
Hence, in order for the Lagrangian in Eq. \eqref{eq:yuk}
to be symmetric under GCP,
we must compare each of these terms with their respective Hermitian conjugate
\begin{equation}
    \begin{split}
        -\mathcal{L}_Y
&\supset \overline{n}^j_R\left[(\Gamma_k)_{ji}
\Phi_k\right]^\dagger q^i_L + \overline{p}^j_R\left[(\Delta_k)_{ji}
\Tilde{\Phi}_k\right]^\dagger q^i_L \\
&= \overline{n}^j_R(\Gamma_k)^*_{ij}
\Phi_k^* q^i_L + \overline{p}^j_R(\Delta_k)^*_{ij}
\Tilde{\Phi}_k^* q^i_L \,\text{.}
    \end{split}
\end{equation}
Thus, the Yukawa Lagrangian in Eq. \eqref{eq:yuk}
is invariant under the GCP transformations
in Eqs. \eqref{GCP_fermion} if and only if
\begin{equation}\label{eq:qrkinv}
    \begin{split}
        \Gamma_b^* &= X_\alpha^\dagger
(X_\theta)_{ab} (\Gamma_a) X_\beta\, , \\
        \Delta_b^* &=  X_\alpha^\dagger
(X_\theta)_{ab}^*(\Delta_a)  X_\gamma \,\text{.}
    \end{split}
\end{equation}
Since we are taking all $X$ matrices to be real,
the condition for the charged $+2/3$ quark matrices $\Delta$ will
yield the same equations as the charged $-1/3$ quark matrices $\Gamma$,
under the substitution $\beta\xrightarrow{}\gamma$.
Therefore, we will focus on the down-type quarks,
and subsequently compute the results for the up-type quarks.

We may write conditions \eqref{eq:qrkinv} as
\ba
&&
X_\alpha \Gamma_1^\ast - (c_\theta \Gamma_1 - s_\theta \Gamma_2) X_\beta = 0,
\nonumber\\
&&
X_\alpha \Gamma_2^\ast - (s_\theta \Gamma_1 + c_\theta \Gamma_2) X_\beta = 0,
\label{GCP_Gamma_12}
\ea
and
\be
X_\alpha \Gamma_3^\ast - \Gamma_3 X_\beta = 0,
\label{GCP_Gamma_3}
\ee
where the basis choices of Eqs.~(\ref{eq:basis}) and (\ref{eq:basis_2})
are implied\footnote{Notice that Eq.~(\ref{GCP_Gamma_3}) is identical to
Eqs.~(\ref{GCP_Gamma_12}) in the limit of $\theta=0$.
}.

Eqs.~(\ref{GCP_Gamma_12}) give us 36 equations in
the 36 unknown real and imaginary parts of the various entries
of the $\Gamma_1$ and $\Gamma_2$ matrices.
In each block we have a system of homogeneous linear equations;
the parameters are zero unless the determinant of the system
vanishes.
Recall that our choice of basis 
decouples the third entry from the other two,
like we did for the scalar potential, 
except that now we may think of 
$\Gamma_k$ as one rank-3 tensor. 
It is for the analysis of the Yukawa matrices that we see the full power
of the new strategy discussed in Section~\ref{sec:new} and
Appendix~\ref{app:torus}.
Next, we turn to a detailed view of this analysis.

\subsection{GCP constraints on Yukawas for down-type quarks}

As mentioned,
our choice of basis for the $X$ matrices decouples the system for
$\Gamma_1$ and $\Gamma_2$ from that of $\Gamma_3$,
and in each we will have 4 distinct regions $mn$, $m3$, $3n$ and $33$.
Thenceforth, we will use the notation $\Gamma_{\bar{k}}$
($\bar{k}=1,2$)
when we wish to refer collectively to $\Gamma_1$ and $\Gamma_2$. 
If we vectorize each block, we can turn Eq. \eqref{eq:qrkinv} into
\begin{equation}
    \begin{split}
        \Gamma_{\{1,2\}}^*&=
        [(R_\theta^\top\otimes R_\alpha^\top \otimes R_\beta^\top)\oplus(R_\theta^\top\otimes R_\alpha^\top)
        \oplus(R_\theta^\top \otimes R_\beta^\top) \oplus R_\theta^\top]\,
        \Gamma_{\{1,2\}}\, , \\
        \Gamma_{\{3\}}^*&=
        [(R_\alpha^\top \otimes R_\beta^\top)\oplus
        (R_\alpha^\top)\oplus
        (R_\beta^\top) \oplus 1]\, \Gamma_{\{3\}}\, ,
    \end{split}
\end{equation}
where
\begin{equation}
    \begin{split}
        \Gamma_{\{1,2\}}=(&(\Gamma_1)_{\{mn\}}, (\Gamma_2)_{\{mn\}},(\Gamma_1)_{\{m3\}},(\Gamma_2)_{\{m3\}},
        (\Gamma_1)_{\{3n\}},(\Gamma_2)_{\{n3\}}, (\Gamma_1)_{33}, (\Gamma_2)_{33})^\top\, ,\\
        \Gamma_{\{3\}}=(&(\Gamma_3)_{\{mn\}},(\Gamma_3)_{\{m3\}},
        (\Gamma_3)_{\{3n\}}, (\Gamma_3)_{33})^\top\, ,
    \end{split}
\end{equation}
using the notation in Eq. \eqref{eq:notation}.

Let us start by looking at the $33$ regions. 
Immediately, we see that
\be
 (\Gamma_3)_{33}^*=(\Gamma_3)_{33}\, ,
\ee
and, thus, $(\Gamma_3)_{33}$ is always real.
As for $(\Gamma_{\bar{k}})_{33}$, we have
\begin{equation}
(\Gamma_{\bar{k}})_{33}^*=R_\theta^\top (\Gamma_{\bar{k}})_{33}^*\ ,\ \ \ \ 
(\Gamma_{\bar{k}})_{33} = ((\Gamma_1)_{33}, (\Gamma_2)_{33})^\top\, .
\end{equation}
Using Eq.~\eqref{3-cases} and the restricted range for $\theta$,
we conclude that either $\theta=0$, in which case $(\Gamma_1)_{33}$
and $(\Gamma_2)_{33}$ are also real,
or, else,
$\theta\neq 0$, in which case $(\Gamma_1)_{33}=(\Gamma_2)_{33}=0$. This constitutes conditions i and ii, respectively, for the $(\Gamma_{\bar{k}})_{33}$ region:

\bgroup
\def\arraystretch{1.5}
\begin{tabular}{rp{15mm}l}
\newtag{i}{k33cond1}. & $\theta=0$ & $(\Gamma_1)_{33}$, $(\Gamma_2)_{33} \in \mathbb{R}\, ,\,$ \\
\newtag{ii}{k33cond2}. & $\theta\neq0$ & $(\Gamma_1)_{33}=(\Gamma_2)_{33}= 0\,.\,$   \\
\end{tabular}
\egroup

\vspace{2mm}

\noindent
The $m3$ ($3n$) region in $\Gamma_3$ is similar under the
substitution
$\theta \xrightarrow{} \alpha$ ($\theta\xrightarrow{}\beta$).
Indeed,
\begin{equation}\label{eq:A9Yuk}
    \begin{split}
        (\Gamma_3)_{\{m3\}}^*=R_\alpha^\top (\Gamma_3)_{\{m3\}}\ ,\ \ \ \ 
        &(\Gamma_3)_{\{m3\}} = ((\Gamma_3)_{13}, (\Gamma_3)_{23})^\top\, , \\
        (\Gamma_3)_{\{3n\}}^*=R_\beta^\top (\Gamma_3)_{\{3n\}}\ ,\ \ \ \ 
        &(\Gamma_3)_{\{3n\}} = ((\Gamma_3)_{31}, (\Gamma_3)_{32})^\top\, .
    \end{split}
\end{equation}
The conditions for the $(\Gamma_{3})_{m3}$ region, then, become

\bgroup
\def\arraystretch{1.5}
\begin{tabular}{rp{15mm}l}
\newtag{i}{3m3cond1}. & $\alpha=0$ & $(\Gamma_3)_{13}$, $(\Gamma_3)_{23} \in \mathbb{R}\, ,\,$ \\
\newtag{ii}{3m3cond2}. & $\alpha\neq0$ & $(\Gamma_3)_{13}=(\Gamma_3)_{23}= 0\,,\,$   \\
\end{tabular}
\egroup

\vspace{2mm}

\noindent
whereas, in the $(\Gamma_{3})_{3n}$ region, we find

\begin{tabular}{rp{15mm}l}
    \renewcommand{\arraystretch}{1.5}
    \newtag{i}{33ncond1}. & $\beta=0$ & $(\Gamma_3)_{31}$, $(\Gamma_3)_{32} \in \mathbb{R}\, ,\,$ \\
    \newtag{ii}{33ncond2}. & $\beta\neq0$ & $(\Gamma_3)_{31}=(\Gamma_3)_{32}= 0\,.$   \\
\end{tabular}

\vspace{2mm}

\noindent
Now looking at systems involving two angles, we have
\begin{equation}\label{eq:A10Yuk}
    \begin{split}
        (\Gamma_{\bar{k}})_{\{m3\}}^*=R_\theta^\top\otimes
R_\alpha^\top(\Gamma_{\bar{k}})_{\{m3\}}^*\ ,\ \ \ 
        &(\Gamma_{\bar{k}})_{\{m3\}} = ((\Gamma_1)_{13}, (\Gamma_1)_{23}, 
        (\Gamma_2)_{13}, (\Gamma_2)_{23})^\top\, , \\
        (\Gamma_{\bar{k}})_{\{3n\}}^*=R_\theta^\top\otimes
R_\beta^\top(\Gamma_{\bar{k}})_{\{3n\}}^*\ ,\ \ \ 
        &(\Gamma_{\bar{k}})_{\{3n\}} = ((\Gamma_1)_{31}, (\Gamma_1)_{32}, 
        (\Gamma_2)_{31}, (\Gamma_2)_{32})^\top\, ,\\
        (\Gamma_3)_{\{mn\}}^*=R_\alpha^\top\otimes R_\beta^\top
        (\Gamma_3)_{\{mn\}}\ ,\ \ \ 
        &(\Gamma_3)_{\{mn\}}=((\Gamma_3)_{11}, (\Gamma_3)_{12}, (\Gamma_3)_{21},(\Gamma_3)_{22})^\top\, .
    \end{split}
\end{equation}
For Eqs. \eqref{eq:A10Yuk}, we shall use the result in Eq.~\eqref{result} for $n=2$,
which simplifies the equations to
\begin{equation}\label{eq:A12Yuk}
    \begin{split}
         \begin{pmatrix}
            (\Gamma_1)_{13} - (\Gamma_2)_{23} \\
            (\Gamma_1)_{23} + (\Gamma_2)_{13}
        \end{pmatrix}^*&=R_{\theta+\alpha}^\top
        \begin{pmatrix}
            (\Gamma_1)_{13} - (\Gamma_2)_{23} \\
            (\Gamma_1)_{23} + (\Gamma_2)_{13}
        \end{pmatrix}\, , \\
        \begin{pmatrix}
            (\Gamma_1)_{13} + (\Gamma_2)_{23} \\
            (\Gamma_1)_{23} - (\Gamma_2)_{13}
        \end{pmatrix}^*&=R_{\alpha-\theta}^\top
        \begin{pmatrix}
            (\Gamma_1)_{13} + (\Gamma_2)_{23} \\
            (\Gamma_1)_{23} - (\Gamma_2)_{13}
        \end{pmatrix}\, ,
    \end{split}
\end{equation}
\begin{equation}\label{eq:A13Yuk}
    \begin{split}
         \begin{pmatrix}
            (\Gamma_1)_{31} - (\Gamma_2)_{32} \\
            (\Gamma_1)_{32} + (\Gamma_2)_{31}
        \end{pmatrix}^*&=R_{\theta+\beta}^\top
        \begin{pmatrix}
            (\Gamma_1)_{31} - (\Gamma_2)_{32} \\
            (\Gamma_1)_{32} + (\Gamma_2)_{31}
        \end{pmatrix}\, , \\
        \begin{pmatrix}
            (\Gamma_1)_{31} + (\Gamma_2)_{32} \\
            (\Gamma_1)_{32} - (\Gamma_2)_{31}
        \end{pmatrix}^*&=R_{\beta-\theta}^\top
        \begin{pmatrix}
            (\Gamma_1)_{31} + (\Gamma_2)_{32} \\
            (\Gamma_1)_{32} - (\Gamma_2)_{31}
        \end{pmatrix}\, ,
    \end{split}
\end{equation}
\begin{equation}\label{eq:A14Yuk}
    \begin{split}
         \begin{pmatrix}
            (\Gamma_3)_{11} - (\Gamma_3)_{22} \\
            (\Gamma_3)_{12} + (\Gamma_3)_{21}
        \end{pmatrix}^*&=R_{\alpha+\beta}^\top
        \begin{pmatrix}
            (\Gamma_3)_{11} - (\Gamma_3)_{22} \\
            (\Gamma_3)_{12} + (\Gamma_3)_{21}
        \end{pmatrix}\, ,\\
        \begin{pmatrix}
            (\Gamma_3)_{11} + (\Gamma_3)_{22} \\
            (\Gamma_3)_{12} - (\Gamma_3)_{21}
        \end{pmatrix}^*&=R_{\beta-\alpha}^\top
        \begin{pmatrix}
            (\Gamma_3)_{11} + (\Gamma_3)_{22} \\
            (\Gamma_3)_{12} - (\Gamma_3)_{21}
        \end{pmatrix}\, .
    \end{split}
\end{equation}
These systems of equations lead to different textures in the cases
where the respective rotation angle is either zero, $\pi$ or any different value.
In Eqs.~\eqref{eq:A12Yuk}, if $\theta+\alpha$ equals zero,
the only possibility is that they are both zero, which means
that $\alpha-\theta$ is also equal to zero in this case.
If instead $\theta+\alpha$ equals $\pi$, the only possibility
is that they both equal $\pi/2$, which again means $\alpha-\theta$ equals zero.
In fact, $\alpha-\theta$ can only be either zero, if $\alpha=\theta$,
or different than zero, if $\alpha\neq\theta$,
but never actually $\pi$.
This means that this region has only 4 possible textures:
$\theta=\alpha=0$; $\theta=\alpha=\pi/2$;
$\theta=\alpha\neq\{0,\pi/2\}$; $\theta\neq\alpha$.
Eqs.~\eqref{eq:A13Yuk} and Eqs.~\eqref{eq:A14Yuk} are
solved in identical fashion.
For the former we have: $\theta=\beta=0$; $\theta=\beta=\pi/2$;
$\theta=\beta\neq\{0,\pi/2\}$; $\theta\neq\beta$.
For the latter we have: $\alpha=\beta=0$; $\alpha=\beta=\pi/2$;
$\alpha=\beta\neq\{0,\pi/2\}$; $\alpha\neq\beta$.

Each condition fully determines the amount of free parameters
in each region.
For the $m3$ region in $\Gamma_{\bar{k}}$, we have:

\bgroup
\def\arraystretch{1.5}
\begin{tabular}{rp{47mm}p{27mm}p{70mm}}
  \newtag{i}{km3cond1}. & $\theta=\alpha=0$& $(\Gamma_{\bar{k}})_{\{m3\}}
\in \mathbb{R}$\, , & \\ 
 \newtag{ii}{km3cond2}. & $\theta=\alpha=\pi/2$& $(\Gamma_{\bar{k}})_{\{m3\}}
\in \mathbb{C}\, ,\,$ & $(\Gamma_1)_{13}= (\Gamma_2)_{23}^*\, ,
\,  (\Gamma_1)_{23}=- (\Gamma_2)_{13}^*$\, , \\ 
\newtag{iii}{km3cond3}. & $\theta=\alpha\neq\{0,\pi/2\} $& $(\Gamma_{\bar{k}})_{\{m3\}} \in \mathbb{R}\, ,\,$ & $(\Gamma_1)_{13}
= (\Gamma_2)_{23}\, ,\,  (\Gamma_1)_{23}
=- (\Gamma_2)_{13}$\, , \\
\newtag{iv}{km3cond4}. & $\theta\neq\alpha$& $(\Gamma_{\bar{k}})_{\{m3\}} = \mathbf{0}$\, . & \\
\end{tabular}
\egroup

\vspace{2mm}

\noindent
Likewise, for the $3n$ region in $\Gamma_{\bar{k}}$,

\bgroup
\def\arraystretch{1.5}
\begin{tabular}{rp{47mm}p{27mm}p{70mm}}
  \newtag{i}{k3ncond1}. & $\theta=\beta=0$ & $(\Gamma_{\bar{k}})_{\{3n\}} \in \mathbb{R}$\, , & \\ 
 \newtag{ii}{k3ncond2}. & $\theta=\beta=\pi/2$& $(\Gamma_{\bar{k}})_{\{3n\}}
\in \mathbb{C}\, ,\,$ & $(\Gamma_1)_{31}
= (\Gamma_2)_{32}^*\, ,\,  (\Gamma_1)_{32}
=- (\Gamma_2)_{31}^* $\, , \\ 
\newtag{iii}{k3ncond3}. & $\theta=\beta\neq\{0,\pi/2\} $& $(\Gamma_{\bar{k}})_{\{3n\}} \in \mathbb{R}\, ,\,$ & $(\Gamma_1)_{31}
= (\Gamma_2)_{32}\, ,\,  (\Gamma_1)_{32}
=- (\Gamma_2)_{31}$\, , \\
\newtag{iv}{k3ncond4}. & $\theta\neq\beta$& $(\Gamma_{\bar{k}})_{\{3n\}} = \mathbf{0}$ \,.& 
\end{tabular}
\egroup

\vspace{2mm}

\noindent
And, the conditions for the $mn$ block in  $\Gamma_3$ are

\bgroup
\def\arraystretch{1.5}
\begin{tabular}{rp{47mm}p{27mm}p{70mm}}
  \newtag{i}{3mncond1}. & $\alpha=\beta=0$ & $(\Gamma_3)_{\{mn\}} \in \mathbb{R}$\, , &  \\ 
 \newtag{ii}{3mncond2}. & $\alpha=\beta=\pi/2$& $(\Gamma_3)_{\{mn\}} \in \mathbb{C}\, ,\,$ & $(\Gamma_3)_{11}
= (\Gamma_3)_{22}^*\, ,\,  (\Gamma_3)_{12}
=- (\Gamma_3)_{21}^* $\, , \\ 
\newtag{iii}{3mncond3}. & $\alpha=\beta\neq\{0,\pi/2\} $& $(\Gamma_3)_{\{mn\}} \in \mathbb{R}\, ,\,$ & $(\Gamma_3)_{11}
= (\Gamma_3)_{22}\, ,\,  (\Gamma_3)_{12}
=- (\Gamma_3)_{21}$\, , \\
\newtag{iv}{3mncond4}. & $\alpha\neq\beta$& $(\Gamma_3)_{\{mn\}} = \mathbf{0}$\, . & 
\end{tabular}
\egroup

\vspace{2mm}

\noindent
Finally, we turn to the
$mn$ sector of $\Gamma_{\bar{k}}$, $\bar{k}=1,2$.
We find
\begin{equation}\label{eq:A11Yuk}
    \begin{split}       
        (\Gamma_{\bar{k}})_{\{mn\}}^*= & R_\theta^\top\otimes R_\alpha^\top\otimes R_\beta^\top
        (\Gamma_{\bar{k}})_{\{mn\}}\, ,\\
        (\Gamma_{\bar{k}})_{\{mn\}}\equiv &
        ((\Gamma_1)_{11}, (\Gamma_1)_{12}, (\Gamma_1)_{21},(\Gamma_1)_{22},
            (\Gamma_2)_{11}, (\Gamma_2)_{12}, (\Gamma_2)_{21},(\Gamma_2)_{22})^\top\, . 
        \end{split}
\end{equation}
For Eqs. \eqref{eq:A11Yuk}, we shall use the result in Eq.~\eqref{result} for $n=3$,
which simplifies the equations to
\begin{equation}\label{eq:A15Yuk}
    \begin{split}
        \begin{pmatrix}
            [(\Gamma_1)_{11} - (\Gamma_1)_{22}] - [(\Gamma_2)_{12} + (\Gamma_2)_{21}] \\
            [(\Gamma_1)_{12} + (\Gamma_1)_{21}] + [(\Gamma_2)_{11} - (\Gamma_2)_{22}]
        \end{pmatrix}^*   &= R_{\theta+\alpha+\beta}^\top \begin{pmatrix}
                                                 [(\Gamma_1)_{11} - (\Gamma_1)_{22}] - [(\Gamma_2)_{12} + (\Gamma_2)_{21}] \\
                                                 [(\Gamma_1)_{12} + (\Gamma_1)_{21}] + [(\Gamma_2)_{11} - (\Gamma_2)_{22}]
                                              \end{pmatrix}\, ,\\
        \begin{pmatrix}
            [(\Gamma_1)_{11} - (\Gamma_1)_{22}] + [(\Gamma_2)_{12} + (\Gamma_2)_{21}] \\
            [(\Gamma_1)_{12} + (\Gamma_1)_{21}] - [(\Gamma_2)_{11} - (\Gamma_2)_{22}]
        \end{pmatrix}^*   &= R_{\alpha+\beta-\theta}^\top \begin{pmatrix}
                                                  [(\Gamma_1)_{11} - (\Gamma_1)_{22}] + [(\Gamma_2)_{12} + (\Gamma_2)_{21}] \\
                                                  [(\Gamma_1)_{12} + (\Gamma_1)_{21}] - [(\Gamma_2)_{11} - (\Gamma_2)_{22}]
                                              \end{pmatrix}\, ,\\
        \begin{pmatrix}
            [(\Gamma_1)_{11} + (\Gamma_1)_{22}] - [(\Gamma_2)_{12} - (\Gamma_2)_{21}] \\
            [(\Gamma_1)_{12} - (\Gamma_1)_{21}] + [(\Gamma_2)_{11} + (\Gamma_2)_{22}]
        \end{pmatrix}^*   &= R_{\theta+\beta-\alpha}^\top \begin{pmatrix}
                                                  [(\Gamma_1)_{11} + (\Gamma_1)_{22}] - [(\Gamma_2)_{12} - (\Gamma_2)_{21}] \\
                                                  [(\Gamma_1)_{12} - (\Gamma_1)_{21}] + [(\Gamma_2)_{11} + (\Gamma_2)_{22}]
                                              \end{pmatrix}\, ,\\
       \begin{pmatrix}
            [(\Gamma_1)_{11} + (\Gamma_1)_{22}] + [(\Gamma_2)_{12} - (\Gamma_2)_{21}] \\
            [(\Gamma_1)_{12} - (\Gamma_1)_{21}] - [(\Gamma_2)_{11} + (\Gamma_2)_{22}]
        \end{pmatrix}^*   &= R_{\theta+\alpha-\beta} \begin{pmatrix}
                                                  [(\Gamma_1)_{11} + (\Gamma_1)_{22}] + [(\Gamma_2)_{12} - (\Gamma_2)_{21}] \\
                                                  [(\Gamma_1)_{12} - (\Gamma_1)_{21}] - [(\Gamma_2)_{11} + (\Gamma_2)_{22}]
                                              \end{pmatrix}\, .\\
    \end{split}
\end{equation}
As before, each angle can either be 0, $\pi$,
or some other value in order to give a unique texture in
the Yukawa matrices.
Naively, there would be a total of $3\times3\times3\times3=81$
different combinations.
But, like before, if for example $\theta+\alpha+\beta=0$,
the only possibility is that $\theta=\alpha=\beta=0$ and
therefore all the angle combinations would also give zero. 
The conditions that the other combinations equal zero are,
respectively, $\alpha+\beta=\theta$, $\theta+\beta=\alpha$,
and $\theta+\alpha=\beta$.
These conditions make the real part of the respective entries
in the Yukawa matrices non-zero.
The conditions for non-zero imaginary parts are,
respectively,
$\theta+\alpha+\beta=\pi$, $\alpha+\beta=\theta+\pi$,
$\theta+\beta=\alpha+\pi$, and $\alpha+\theta=\beta+\pi$.
However,
the last three can only be achieved if the two angles on
the left-hand side equal $\pi/2$ and the other one equals zero,
which means that $\theta+\alpha+\beta$ also equals $\pi$.
Performing all valid combinations results in just 15 different textures for the $mn$ system in $\Gamma_1$ and $\Gamma_2$:

\bgroup
\def\arraystretch{1.5}
\begin{longtable}{rp{47mm}p{27mm}p{70mm}}
  \newtag{i}{kmncond1}. & $\theta=\alpha=\beta=0$ &
$(\Gamma_{\bar k})_{\{mn\}} \in \mathbb{R},$ &  \\
  \newtag{ii}{kmncond2}. & $\beta=0$, $\theta=\alpha=\pi/2$
& $(\Gamma_{\bar k})_{\{mn\}} \in \mathbb{C},$ &  $(\Gamma_1)_{11}=(\Gamma_2)_{21}^*,
    (\Gamma_1)_{12}=(\Gamma_2)_{22}^*,$ \newline
    $(\Gamma_1)_{21}=-(\Gamma_2)_{11}^*,
    (\Gamma_1)_{22}=-(\Gamma_2)_{12}^* $ \\
  \newtag{iii}{kmncond3}. & $\beta=0$, $\theta=\alpha\neq\{0,\pi/2\}$
& $(\Gamma_{\bar k})_{\{mn\}} \in \mathbb{R},$ &  $(\Gamma_1)_{11}=(\Gamma_2)_{21},
    (\Gamma_1)_{12}=(\Gamma_2)_{22},$ \newline
    $(\Gamma_1)_{21}=-(\Gamma_2)_{11},
    (\Gamma_1)_{22}=-(\Gamma_2)_{12} $ \\ 
   \newtag{iv}{kmncond4}. & $\alpha=0$, $\theta=\beta=\pi/2$ & 
    $(\Gamma_{\bar k})_{\{mn\}} \in \mathbb{C},$ &$ 
    (\Gamma_1)_{11}=(\Gamma_2)_{12}^*,
    (\Gamma_1)_{12}=-(\Gamma_2)_{11}^*,$ \newline
    $(\Gamma_1)_{21}=(\Gamma_2)_{22}^*,
    (\Gamma_1)_{22}=-(\Gamma_2)_{21}^* $ \\
    \newtag{v}{kmncond5}. & $\alpha=0$, $\theta=\beta\neq\{0,\pi/2\}$ &
$(\Gamma_{\bar k})_{\{mn\}} \in \mathbb{R},$ & $ 
    (\Gamma_1)_{11}=(\Gamma_2)_{12},
    (\Gamma_1)_{12}=-(\Gamma_2)_{11},$ \newline
    $(\Gamma_1)_{21}=(\Gamma_2)_{22},
    (\Gamma_1)_{22}=-(\Gamma_2)_{21}, $ \\
    \newtag{vi}{kmncond6}. & $\theta=0$, $\alpha=\beta=\pi/2$ &
$(\Gamma_{\bar k})_{\{mn\}} \in \mathbb{C},$ & $(\Gamma_1)_{11}=(\Gamma_1)_{22}^*,
    (\Gamma_1)_{12}=-(\Gamma_1)_{21}^*,$ \newline $(\Gamma_2)_{11}=(\Gamma_2)_{22}^*,
    (\Gamma_2)_{12}=-(\Gamma_2)_{21}^*, $\\
    \newtag{vii}{kmncond7}. & $\theta=0$, $\alpha=\beta\neq\{0,\pi/2\}$ &
$(\Gamma_{\bar k})_{\{mn\}} \in \mathbb{R},$ & $(\Gamma_1)_{11}=(\Gamma_1)_{22},
    (\Gamma_1)_{12}=-(\Gamma_1)_{21},$ \newline $(\Gamma_2)_{11}=(\Gamma_2)_{22},
    (\Gamma_2)_{12}=-(\Gamma_2)_{21}, $\\
    \newtag{viii}{kmncond8}. & $\theta=\pi/2$ , $\alpha+\beta=\pi/2$,\newline
    $\alpha,\beta\neq\{0,\pi/2\}$ & $(\Gamma_{\bar k})_{\{mn\}} \in \mathbb{C},$ & $(\Gamma_1)_{11}=-(\Gamma_1)_{22}=(\Gamma_2)_{12}^*=(\Gamma_2)_{21}^*,$ \newline $(\Gamma_1)_{12}=(\Gamma_1)_{21}=-(\Gamma_2)_{11}^*=(\Gamma_2)_{22}^*, $\\
    \newtag{ix}{kmncond9}. & $\alpha=\pi/2$,
    $\theta+\beta=\pi/2$,\newline
    $\theta,\beta\neq\{0,\pi/2\}$ & $(\Gamma_{\bar k})_{\{mn\}} \in \mathbb{C},$ & $(\Gamma_1)_{11}=-(\Gamma_2)_{12}=(\Gamma_1)_{22}^*=(\Gamma_2)_{21}^*,$ \newline $(\Gamma_1)_{12}=(\Gamma_2)_{11}=-(\Gamma_1)_{21}^*=(\Gamma_2)_{22}^*,$\\
    \newtag{x}{kmncond10}. & $\beta=\pi/2$,
    $\theta+\alpha=\pi/2$,\newline
    $\theta,\alpha\neq\{0,\pi/2\}$ & $(\Gamma_{\bar k})_{\{mn\}} \in \mathbb{C},$ & $(\Gamma_1)_{11}=-(\Gamma_2)_{21}=(\Gamma_1)_{22}^*=(\Gamma_2)_{12}^*,$ \newline $ -(\Gamma_1)_{12}=(\Gamma_2)_{22}=(\Gamma_1)_{21}^*=(\Gamma_2)_{11}^*, $\\
    \newtag{xi}{kmncond11}. &
$\alpha+\beta=\theta$, \newline $\theta\neq\{\alpha,\beta,\pi/2\}$
& $(\Gamma_{\bar k})_{\{mn\}} \in \mathbb{R}, $ &
$ (\Gamma_1)_{11}=-(\Gamma_1)_{22}=(\Gamma_2)_{12}=(\Gamma_2)_{21},
$ \newline $(\Gamma_1)_{12}=(\Gamma_1)_{21}=-(\Gamma_2)_{11}=(\Gamma_2)_{22},$\\
    \newtag{xii}{kmncond12}. & $\theta+\beta=\alpha$, \newline  $\alpha\neq\{\theta,\beta,\pi/2\}$ & $(\Gamma_{\bar k})_{\{mn\}} \in \mathbb{R}, $ & $ (\Gamma_1)_{11}=(\Gamma_1)_{22}=-(\Gamma_2)_{12}=(\Gamma_2)_{21},$ \newline $(\Gamma_1)_{12}=-(\Gamma_1)_{21}=(\Gamma_2)_{11}=(\Gamma_2)_{22}, $\\
    \newtag{xiii}{kmncond13}. &  $\theta+\alpha=\beta$, \newline  $\beta\neq\{\theta,\alpha,\pi/2\}$ & $(\Gamma_{\bar k})_{\{mn\}} \in \mathbb{R}, $ & $ (\Gamma_1)_{11}=(\Gamma_1)_{22}=(\Gamma_2)_{12}=-(\Gamma_2)_{21},$ \newline $-(\Gamma_1)_{12}=(\Gamma_1)_{21}=(\Gamma_2)_{11}=(\Gamma_2)_{22},$\\
    \newtag{xiv}{kmncond14}. &  $\theta+\alpha+\beta=\pi$, \newline  $\theta,\alpha,\beta\neq\pi/2$ & $(\Gamma_{\bar k})_{\{mn\}} \in \mathbb{I},$ & $ -(\Gamma_1)_{11}=(\Gamma_1)_{22}=(\Gamma_2)_{12}=(\Gamma_2)_{21},$ \newline $(\Gamma_1)_{12}=(\Gamma_1)_{21}=(\Gamma_2)_{11}=-(\Gamma_2)_{22}, $\\
    \newtag{xv}{kmncond15}. &  $\alpha+\beta\neq\theta$,
    $\theta+\beta\neq\alpha$, \newline  $\theta+\alpha\neq\beta$, 
    $\theta+\alpha+\beta\neq\pi$ & $(\Gamma_{\bar k})_{\{mn\}}=\mathbf{0}$. & \\

\end{longtable}
\egroup

The next step is to combine these 15 conditions with the 18
from the other blocks and then we should have all different
combinations of angles $\theta$, $\alpha$ and $\beta$ that
produce unique textures on $\Gamma_k$, and thus on $M_d$. 
Recall that, in the spirit that adding to the scalar potential
soft breaking terms might be useful for some model building,
we are using the most general vevs $v_i$.

Excluding combinations that lead either to
null or degenerate eigenvalues,
using the notation $(\theta, \alpha, \beta)$,
we have:
\setcounter{table}{0}
\bgroup
\def\arraystretch{1.5}
\begin{longtable}{rl|c|c|c|c|c|c|c}
   &  & $(\Gamma_{\bar{k}})_{33}$ & $(\Gamma_{3})_{m3}$ & $(\Gamma_{3})_{3n}$ & $(\Gamma_{\bar{k}})_{m3}$ & $(\Gamma_{\bar{k}})_{3n}$ & $(\Gamma_{3})_{mn}$ & $(\Gamma_{\bar{k}})_{mn}$ \\ \hline
    \endfirsthead
    &  & $(\Gamma_{\bar{k}})_{33}$ & $(\Gamma_{3})_{m3}$ & $(\Gamma_{3})_{3n}$ & $(\Gamma_{\bar{k}})_{m3}$ & $(\Gamma_{\bar{k}})_{3n}$ & $(\Gamma_{3})_{mn}$ & $(\Gamma_{\bar{k}})_{mn}$ \\
    \hline
    \endhead
    \hline
    \endfoot
    \endlastfoot

\newtag{1}{text1}. & $(0, 0, 0)$ & \ref{k33cond1} & \ref{3m3cond1} & \ref{33ncond1} & \ref{km3cond1} & \ref{k3ncond1} & \ref{3mncond1} & \ref{kmncond1} \\ \hline
\newtag{2}{text2}. & $(\pi/2, \pi/2, 0)$ & \ref{k33cond2} & \ref{3m3cond2} & \ref{33ncond1} & \ref{km3cond2} & \ref{k3ncond4} & \ref{3mncond4} & \ref{kmncond2} \\ \hline
\newtag{3}{text3}. & $(\theta, \theta, 0)$ & \ref{k33cond2} & \ref{3m3cond2} & \ref{33ncond1} & \ref{km3cond3} & \ref{k3ncond4} & \ref{3mncond4} & \ref{kmncond3} \\ \hline
\newtag{4}{text4}. & $(\pi/2, 0, \pi/2)$ & \ref{k33cond2} & \ref{3m3cond1} & \ref{33ncond2} & \ref{km3cond4} & \ref{k3ncond2} & \ref{3mncond4} & \ref{kmncond4} \\ \hline
\newtag{5}{text5}. & $(\theta, 0, \theta)$ & \ref{k33cond2} & \ref{3m3cond1} & \ref{33ncond2} & \ref{km3cond4} & \ref{k3ncond3} & \ref{3mncond4} & \ref{kmncond5} \\ \hline
\newtag{6}{text6}. & $(0, \pi/2, \pi/2)^*$ & \ref{k33cond1} & \ref{3m3cond2} & \ref{33ncond2} & \ref{km3cond4} & \ref{k3ncond4} & \ref{3mncond2} & \ref{kmncond6} \\ \hline
\newtag{7}{text7}. & $(0, \alpha, \alpha)^*$ & \ref{k33cond1} & \ref{3m3cond2} & \ref{33ncond2} & \ref{km3cond4} & \ref{k3ncond4} & \ref{3mncond3} & \ref{kmncond7} \\ \hline
\newtag{8}{text8}. & $(\pi/2, \pi/4, \pi/4)^*$ & \ref{k33cond2} & \ref{3m3cond2} & \ref{33ncond2} & \ref{km3cond4} & \ref{k3ncond4} & \ref{3mncond3} & \ref{kmncond8} \\ \hline
\newtag{9}{text9}. & $(\pi/2, \alpha, \pi/2 - \alpha)^*$ &  \ref{k33cond2} & \ref{3m3cond2} & \ref{33ncond2} & \ref{km3cond4} & \ref{k3ncond4} & \ref{3mncond4} & \ref{kmncond8} \\ \hline
\newtag{10}{text10}. & $(\pi/4, \pi/2, \pi/4)$ & \ref{k33cond2} & \ref{3m3cond2} & \ref{33ncond2} & \ref{km3cond4} & \ref{k3ncond4} & \ref{3mncond4} & \ref{kmncond9} \\ \hline
\newtag{11}{text11}. & $(\theta, \pi/2, \pi/2 - \theta)^*$ &  \ref{k33cond2} & \ref{3m3cond2} & \ref{33ncond2} & \ref{km3cond4} & \ref{k3ncond4} & \ref{3mncond4} & \ref{kmncond9} \\ \hline
\newtag{12}{text12}. & $(\pi/4, \pi/4, \pi/2)$ & \ref{k33cond2} & \ref{3m3cond2} & \ref{33ncond2} & \ref{km3cond3} & \ref{k3ncond4} & \ref{3mncond4} & \ref{kmncond10} \\ \hline
\newtag{13}{text13}. & $(\theta, \pi/2 - \theta, \pi/2)^*$ &  \ref{k33cond2} & \ref{3m3cond2} & \ref{33ncond2} & \ref{km3cond4} & \ref{k3ncond4} & \ref{3mncond4} & \ref{kmncond10} \\ \hline
\newtag{14}{text14}. & $(\theta, \theta/2, \theta/2)^*$ & \ref{k33cond2} & \ref{3m3cond2} & \ref{33ncond2} & \ref{km3cond4} & \ref{k3ncond4} & \ref{3mncond3}  & \ref{kmncond11} \\ \hline
\newtag{15}{text15}. & $(\theta, \alpha, \theta - \alpha)^*$ & \ref{k33cond2} & \ref{3m3cond2}  & \ref{33ncond2} & \ref{km3cond4} & \ref{k3ncond4} & \ref{3mncond4} &  \ref{kmncond11}\\ \hline
\newtag{16}{text16}. & $(\theta, 2\theta, \theta)$ & \ref{k33cond2} & \ref{3m3cond2} & \ref{33ncond2} & \ref{km3cond4} & \ref{k3ncond3} & \ref{3mncond4} & \ref{kmncond12} \\ \hline
\newtag{17}{text17}. & $(\theta, \alpha, \alpha - \theta)^*$ & \ref{k33cond2} & \ref{3m3cond2} & \ref{33ncond2} & \ref{km3cond4} & \ref{k3ncond4} & \ref{3mncond4} & \ref{kmncond12} \\ \hline
\newtag{18}{text18}. & $(\theta, \theta, 2\theta)$ & \ref{k33cond2} & \ref{3m3cond2} & \ref{33ncond2} & \ref{km3cond3} & \ref{k3ncond4} & \ref{3mncond4} & \ref{kmncond13} \\ \hline
\newtag{19}{text19}. & $(\theta, \alpha, \theta + \alpha)^*$ & \ref{k33cond2} & \ref{3m3cond2} & \ref{33ncond2} & \ref{km3cond4} & \ref{k3ncond4} & \ref{3mncond4} & \ref{kmncond13} \\ \hline
\newtag{20}{text20}. & $(\pi/3, \pi/3, \pi/3)$ & \ref{k33cond2} & \ref{3m3cond2} & \ref{33ncond2} & \ref{km3cond3} & \ref{k3ncond3} & \ref{3mncond3}  & \ref{kmncond14} \\ \hline
\newtag{21}{text21}. & $(\theta, \theta, \pi - 2\theta)$ &  \ref{k33cond2} & \ref{3m3cond2} & \ref{33ncond2} & \ref{km3cond3} & \ref{k3ncond4} & \ref{3mncond4} &  \ref{kmncond14}\\ \hline
\newtag{22}{text22}. & $(\theta, \pi - 2\theta, \theta)$ &  \ref{k33cond2} & \ref{3m3cond2} & \ref{33ncond2} & \ref{km3cond4} & \ref{k3ncond3} & \ref{3mncond4} &  \ref{kmncond14}\\ \hline
\newtag{23}{text23}. & $(\theta, (\pi - \theta)/2, (\pi - \theta)/2)^*$ & \ref{k33cond2} &  \ref{3m3cond2} & \ref{33ncond2} & \ref{km3cond4} & \ref{k3ncond4} & \ref{3mncond3} & \ref{kmncond14} \\ \hline
\newtag{24}{text24}. & $(\theta, \alpha, \pi - \theta - \alpha)^*$ & \ref{k33cond2} & \ref{3m3cond2} & \ref{33ncond2} & \ref{km3cond4} & \ref{k3ncond4} & \ref{3mncond4} & \ref{kmncond14} \\ \hline
\newtag{25}{text25}. & $(\theta, 0, 0)^\diamond$ & \ref{k33cond2} & \ref{3m3cond1} & \ref{33ncond1} & \ref{km3cond4} & \ref{k3ncond4} & \ref{3mncond1} & \ref{kmncond15} \\ \hline
\newtag{26}{text26}. & $(\pi/2, \pi/2, \pi/2)$ & \ref{k33cond2} & \ref{3m3cond2} & \ref{33ncond2} & \ref{km3cond2} & \ref{k3ncond2} & \ref{3mncond2} & \ref{kmncond15} \\ \hline
\newtag{27}{text27}. & $(\theta, \theta, \theta)$ & \ref{k33cond2} & \ref{3m3cond2} & \ref{33ncond2} & \ref{km3cond3} & \ref{k3ncond3}  & \ref{3mncond3} & \ref{kmncond15} \\

\caption{\label{tab:cases_condsYuk} Combinations of $(\theta, \alpha, \beta)$ and respective conditions for the 33, m3, 3n, and mn blocks of $\Gamma_{\bar{k}}$ and $\Gamma_{3}$.}
\\
\end{longtable}
\egroup

An asterisk here means that the third quark decouples,
i.e. $M_d$ is block diagonal.
A diamond here means that the quarks couple only
to $\Phi_3$, i.e. $\Gamma_1=\Gamma_2=\mathbf{0}$.

\subsection{GCP constraints on the whole Yukawa sector}

Now, we must combine the cases from the down-type $\Gamma_k$
matrices with the ones from the up-type $\Delta_k$ matrices,
which, for our choice of basis, are the same under the replacement
$\beta\xrightarrow{}\gamma$.
The final cases $(\theta,\alpha,\beta,\gamma)$ must have the
same value of $\theta$ and $\alpha$ for the up and
down matrices simultaneously, 
and we must not combine two cases bearing an asterisk,
as that would lead to a block-diagonal CKM matrix,
which is ruled out experimentally.
We exclude those textures leading to a massless quark
or to a block-diagonal CKM matrix.
Thus, we have:

\begin{longtable}{|>{\centering\arraybackslash}m{5cm}|>{\centering\arraybackslash}m{2cm}|>{\centering\arraybackslash}m{2cm}|}
    \hline
    $(\theta,\alpha,\beta,\gamma)$ & $(\theta,\alpha,\beta)$ & $(\theta,\alpha,\gamma)$ \\
    \hline
    \endfirsthead
    \hline
    $(\theta,\alpha,\beta,\gamma)$ & $(\theta,\alpha,\beta)$ & $(\theta,\alpha,\gamma)$ \\
    \hline
    \endhead
    \hline
    
    \endfoot
    \endlastfoot

    $(0,0,0,0)$ & \ref{text1} & \ref{text1} \\
    $(\pi/2,\pi/2,0,0)$ & \ref{text2} & \ref{text2} \\
    $(\pi/2,\pi/2,0,\pi/2)$ & \ref{text2} & \ref{text26} \\
    $(\theta,\theta,0,0)$ & \ref{text3} & \ref{text3} \\
    $(\pi/4,\pi/4,0,\pi/2)$ & \ref{text3} & \ref{text12} \\
    $(\theta,\theta,0, 2\theta)$ & \ref{text3} & \ref{text18} \\
    $(\pi/3,\pi/3,0, \pi/3)$ & \ref{text3} & \ref{text20} \\
    $(\theta,\theta,0, \pi-2\theta)$ & \ref{text3} & \ref{text21} \\
    $(\theta,\theta,0, \theta)$ & \ref{text3} & \ref{text27} \\
    $(\pi/2,0, \pi/2,\pi/2)$ & \ref{text4} & \ref{text4} \\
    $(\pi/2,0, \pi/2,0)$ & \ref{text4} & \ref{text25} \\
    $(\theta,0, \theta,\theta)$ & \ref{text5} & \ref{text5} \\
    $(\theta,0, \theta,0)$ & \ref{text5} & \ref{text25} \\
    $(\pi/4,\pi/2,\pi/4,\pi/4)$ & \ref{text10} & \ref{text10} \\
    $(\pi/4,\pi/4,\pi/2,0)$ & \ref{text12} & \ref{text3} \\
    $(\pi/4,\pi/4,\pi/2,\pi/2)$ & \ref{text12} & \ref{text12} \\
    $(\pi/4,\pi/4,\pi/2,\pi/4)$ & \ref{text12} & \ref{text27} \\
    $(\pi/6,\pi/3, \pi/2,\pi/6)$ & \ref{text13} & \ref{text16} \\
    $(\frac{2\pi}{5},\pi/5, \pi/5,\frac{2\pi}{5})$ & \ref{text14} & \ref{text22} \\
    $(\theta,\pi-2\theta, 3\theta-\pi,\theta)$ & \ref{text15} & \ref{text22} \\
    $(\pi/6,\pi/3, \pi/6,\pi/2)$ & \ref{text16} & \ref{text13} \\
    $(\theta,2\theta,\theta,\theta)$ & \ref{text16} & \ref{text16} \\
    $(\theta,2\theta,\theta,3\theta)$ & \ref{text16} & \ref{text19} \\
    $(\pi/5,\frac{2\pi}{5},\pi/5,\frac{2\pi}{5})$ & \ref{text16} & \ref{text23} \\
    $(\theta,2\theta,\theta,\pi-3\theta)$ & \ref{text16} & \ref{text24} \\
    $(\theta,\pi-2\theta,\pi-3\theta,\theta)$ & \ref{text17} & \ref{text22} \\
    $(\theta,\theta,2\theta,0)$ & \ref{text18} & \ref{text3} \\
    $(\theta,\theta,2\theta,2\theta)$ & \ref{text18} & \ref{text18} \\
    $(\theta,\theta,2\theta,\theta)$ & \ref{text18} & \ref{text27} \\
    $(\theta,2\theta,3\theta,\theta)$ & \ref{text19} & \ref{text16} \\
    $(\pi/3,\pi/3, \pi/3,0)$ & \ref{text20} & \ref{text3} \\
    $(\pi/3,\pi/3, \pi/3,\pi/3)$ & \ref{text20} & \ref{text20} \\
    $(\theta,\theta, \pi-2\theta,0)$ & \ref{text21} & \ref{text3} \\
    $(\theta,\theta, \pi-2\theta,\pi-2\theta)$ & \ref{text21} & \ref{text21} \\
    $(\theta,\theta, \pi-2\theta,\theta)$ & \ref{text21} & \ref{text27} \\
    $(\frac{2\pi}{5},\pi/5,\frac{2\pi}{5}, \pi/5)$ & \ref{text22} & \ref{text14} \\
    $(\theta,\pi-2\theta,\theta, 3\theta-\pi)$ & \ref{text22} & \ref{text15} \\
    $(\theta,\pi-2\theta,\theta, \pi-3\theta)$& \ref{text22} & \ref{text17} \\
    $(\theta,\pi-2\theta,\theta, \theta)$ & \ref{text22} & \ref{text22} \\
    $(\pi/5,\frac{2\pi}{5},\frac{2\pi}{5},\pi/5)$ & \ref{text23} & \ref{text16} \\
    $(\theta,2\theta,\pi-3\theta,\theta)$ & \ref{text24} & \ref{text16} \\
    $(\pi/2,0,0, \pi/2)$ & \ref{text25} & \ref{text4} \\
    $(\theta,0,0, \theta)$ & \ref{text25} & \ref{text5} \\
    $(\theta,0,0, 0)$ & \ref{text25} & \ref{text25} \\
    $(\pi/2,\pi/2,\pi/2,0)$ & \ref{text26} & \ref{text2} \\
    $(\pi/2,\pi/2,\pi/2,\pi/2)$ & \ref{text26} & \ref{text26} \\
    $(\theta,\theta, \theta,0)$ & \ref{text27} & \ref{text3} \\
    $(\pi/4,\pi/4,\pi/4, \pi/2)$ & \ref{text27} & \ref{text12} \\
    $(\theta,\theta, \theta,2\theta)$ & \ref{text27} & \ref{text18} \\ 
    $(\theta,\theta, \theta,\pi-2\theta)$ & \ref{text27} & \ref{text21} \\
    $(\theta,\theta, \theta,\theta)$ & \ref{text27} & \ref{text27} \\\hline
    
\caption{Combinations of $(\theta, \alpha, \beta, \gamma)$ and respective constraints on the up and down Yukawa sectors.}
\label{tab:cases_tabg}\\
\end{longtable}

\vspace{4ex}

These 51 combinations of angles $(\theta,\alpha,\beta,\gamma)$
for the GCP transformations are all the combinations
that lead to unique textures on the mass matrices
$M_u$ and $M_d$, consistent with non-zero and non-degenerate
quark masses and a CKM matrix which is not block-diagonal.

But there is one further simple constraint, regarding CP violation.
Experiments in $B$ decays prompt the conclusion that there must
be CP violation in the CKM matrix, which is related to a physical
quantity called the Jarlskog invariant $J_{CP}$
\cite{Jarlskog:1985ht,Jarlskog:1985cw}.
In terms of the CKM matrix elements this invariant is
given by \cite{ParticleDataGroup:2022pth}
\begin{equation}
    J_{CP} = \text{Im}(V_{us}V_{cb}V^*_{ub}V^*_{cs})
= (3.08^{+0.15}_{-0.13})\times 10^{-5}\,\text{.}
\end{equation}
One can construct another related CP-violating, basis invariant, quantity
\cite{Bernabeu:1986fc,Jarlskog:1985ht,Jarlskog:1985cw}
\begin{equation}
    \begin{split}
        J = \text{Tr}\left[ H_u, H_d\right]^3 
        &= 6i(m_t^2-m_c^2)(m_t^2-m_u^2)(m_c^2-m_u^2)\\
        &\times (m_b^2-m_s^2)(m_b^2-m_d^2)(m_s^2-m_d^2) J_{CP}\,\text{.}
    \end{split}
\end{equation}
We have already excluded models with quarks bearing zero mass
or equal mass among each other.
Therefore, $J=0$ if and only
if the CKM matrix conserves CP, i.e. $J_{CP}=0$,
which is ruled out experimentally. 

If we compute $H_u$, $H_d$ and $J$ for each of our 51 models, we find that for
$(\pi/6,\pi/3,\pi/2,\pi/6)$, 
$(\theta,\pi-2\theta,3\theta-\pi,\theta)$,
$(\pi/6,\pi/3,\pi/6,\pi/2)$,
$(\theta,2\theta,\theta,3\theta)$,
$(\theta,2\theta,\theta,\pi-3\theta)$,
$(\theta,\pi-2\theta,\pi-3\theta,\theta)$,
$(\theta,2\theta,3\theta,\theta)$,
$(\theta,\pi-2\theta,\theta ,3\theta-\pi)$,
$(\theta,\pi-2\theta,\theta,\pi-3\theta)$,
$(\theta,2\theta,\pi-3\theta,\theta)$ and
$(\theta,0,0,0)$,
$J$ is always zero.
Therefore, we can eliminate these models from our list,
leaving us with just 40 models.

Finally, we can list every GCP symmetry that can be imposed
on the 3HDM consistent with non-zero and non-degenerate
quark masses and with non-diagonal and CP-violating CKM matrix.
For each model we note the allowed values for $\theta$,
from which we can deduce the form of the scalar potential
(CPa, CPb, CPc, or CPd),
along with the amount of real independent parameters in
the down-type and up-type Yukawa matrices.
This is shown in Table \ref{tab:final_cases}.
An asterisk means that, for that Yukawa matrix,
the third quark decouples from the first two and a diamond
means that the quarks couple only to $\Phi_3$. 

\vspace{8mm}

\begin{longtable}{>{\centering\arraybackslash}m{3.5cm}|>{\centering\arraybackslash}m{2.5cm}|>{\centering\arraybackslash}m{2.8cm}|>{\centering\arraybackslash}m{2.8cm}}
\hline
\hline
    $(\theta,\alpha,\beta,\gamma)$ & Range for $\theta$ & \# of real param. in down-type Yukawa & \# of real param. in up-type Yukawa\\
\hline
\endfirsthead

\hline
    $(\theta,\alpha,\beta,\gamma)$ & Range for $\theta$ & \# of real param. in down-type Yukawa & \# of real param. in up-type Yukawa\\
\hline
\endhead

\hline
\endfoot

\endlastfoot

 & & & \\[-1em]
$(0,0,0,0)$ & 0 & 27 & 27\\[0.1em]
\hline

 & & & \\[-1em]
$(\frac{\pi}{2},0,0,\frac{\pi}{2})$ & \multirow{7}{*}{$\frac{\pi}{2}$} & $9^\diamond$ & 15 \\[0.1em]
$(\frac{\pi}{2},0,\frac{\pi}{2},0)$ &  & 15 & $9^\diamond$\\[0.1em]
$(\frac{\pi}{2},0,\frac{\pi}{2},\frac{\pi}{2})$ &  & 15 & 15\\[0.1em]
$(\frac{\pi}{2},\frac{\pi}{2},0,0)$ &  & 15 & 15\\[0.1em]
$(\frac{\pi}{2},\frac{\pi}{2},0,\frac{\pi}{2})$ &  & 15 & 13\\[0.1em]
$(\frac{\pi}{2},\frac{\pi}{2},\frac{\pi}{2},0)$ &   & 13& 15\\[0.1em]
$(\frac{\pi}{2},\frac{\pi}{2},\frac{\pi}{2},\frac{\pi}{2})$ &   & 13& 13\\[0.2em]
\hline

 & & & \\[-1em]
$(\frac{\pi}{3},\frac{\pi}{3},0,\frac{\pi}{3})$ &  \multirow{3}{*}{$\frac{\pi}{3}$} & 9 & 9\\[0.1em]
$(\frac{\pi}{3},\frac{\pi}{3},\frac{\pi}{3},0)$ &   & 9 & 9\\[0.1em]
$(\frac{\pi}{3},\frac{\pi}{3},\frac{\pi}{3},\frac{\pi}{3})$ &   & 9 & 9\\[0.2em]
\hline

 & & & \\[-1em]
$(\frac{\pi}{4},\frac{\pi}{4},0,\frac{\pi}{2})$ &  \multirow{6}{*}{$\frac{\pi}{4}$} & 9 & 7\\[0.1em]
$(\frac{\pi}{4},\frac{\pi}{4},\frac{\pi}{2},0)$ &   & 7 & 9\\[0.1em]
$(\frac{\pi}{4},\frac{\pi}{4},\frac{\pi}{4},\frac{\pi}{2})$ &  & 7 & 7\\[0.1em]
$(\frac{\pi}{4},\frac{\pi}{4},\frac{\pi}{2},\frac{\pi}{4})$ &   & 7 & 7\\[0.1em]
$(\frac{\pi}{4},\frac{\pi}{4},\frac{\pi}{2},\frac{\pi}{2})$ &   & 7 & 7\\[0.1em]
$(\frac{\pi}{4},\frac{\pi}{2},\frac{\pi}{4},\frac{\pi}{4})$ &   & 7 & 7\\[0.2em]
\hline 

 & & & \\[-1em]
$(\frac{\pi}{5},\frac{2\pi}{5},\frac{\pi}{5},\frac{2\pi}{5})$ &  \multirow{2}{*}{$\frac{\pi}{5}$} & 5 & $5^*$\\[0.1em]
$(\frac{\pi}{5},\frac{2\pi}{5},\frac{2\pi}{5},\frac{\pi}{5})$ & & $5^*$ & 5\\[0.2em]
\hline

 & & & \\[-1em]
$(\frac{2\pi}{5},\frac{\pi}{5},\frac{2\pi}{5},\frac{\pi}{5})$ &  \multirow{2}{*}{$\frac{2\pi}{5}$} & 5 & $5^*$\\[0.1em]
$(\frac{2\pi}{5},\frac{\pi}{5},\frac{\pi}{5},\frac{2\pi}{5})$ & & $5^*$ & 5\\[0.2em]
\hline \pagebreak

 & & & \\[-1em]
$(\theta,0,0,\theta)$ & \multirow{4}{*}{$\left(0, \frac{\pi}{2}\right)$}  & $9^\diamond$ & $9$\\[0.1em]
$(\theta,0,\theta,0)$ &   & $9$& $9^\diamond$\\[0.1em]
$(\theta,0,\theta,\theta)$ &   & $9$& $9$\\[0.1em]
$(\theta,\theta,0,0)$ &   & $9$& $9$\\[0.2em]
\hline

 & & & \\[-1em]
$(\theta,\theta,0,\theta)$ & \multirow{3}{*}{$\left(0, \frac{\pi}{2}\right)\setminus\{\frac{\pi}{3}\}$}   & $9$ & $7$\\[0.1em]
$(\theta,\theta,\theta,0)$ &  & $7$ & $9$\\[0.1em]
$(\theta,\theta,\theta,\theta)$ &   & $7$& $7$\\[0.2em]
\hline

 & & & \\[-1em]
$(\theta,\theta,0,2\theta)$ &  \multirow{6}{*}{ $\left(0, \frac{\pi}{4}\right)$} & $9$ & $5$\\[0.1em]
$(\theta,\theta,2\theta,0)$ &  & $5$ & $9$\\[0.1em]
$(\theta,\theta,\theta,2\theta)$ & & $7$ & $5$\\[0.1em]
$(\theta,\theta,2\theta,\theta)$ &  & $5$ & $7$\\[0.1em]
$(\theta,\theta,2\theta,2\theta)$ &  & $5$ & $5$\\[0.1em]
$(\theta,2\theta,\theta,\theta)$ &  & $5$ & $5$\\[0.2em]
\hline

 & & & \\[-1em]
$(\theta,\theta,0,\pi-2\theta)$ &  \multirow{6}{*}{ $\left(\frac{\pi}{4}, \frac{\pi}{2}\right)\setminus\{\frac{\pi}{3}\}$}  & $9$ & $5$\\[0.1em]
$(\theta,\theta,\pi-2\theta,0)$ &  & $5$ & $9$\\[0.1em]
$(\theta,\theta,\theta,\pi-2\theta)$ & & $7$ & $5$\\[0.1em]
$(\theta,\theta,\pi-2\theta,\theta)$ &  & $5$ & $7$\\[0.1em]
$(\theta,\theta,\pi-2\theta,\pi-2\theta)$ &  & $5$ & $5$\\[0.1em]
$(\theta,\pi-2\theta,\theta,\theta)$ &  & $5$ & $5$\\[0.2em] 
\hline
\hline
\caption{\label{tab:final_cases}All 40 GCP-symmetric 3HDM compatible with non zero and non-degenerate quark masses as well as a non-vanishing Jarlskog invariant.}
\end{longtable}

Table \ref{tab:final_cases} is one main result from our work.
Notice the dramatic reduction in the number of unknowns. Indeed,
in each charge sector we would generally have 3 matrices,
with 9 complex entries for a total of 54 (real) parameters.
Imposing the various GCP symmetries reduces these numbers to those
shown in Table \ref{tab:final_cases}.

\section{\label{sec:yukawa_matrices}Yukawa textures consistent with each GCP implementation}

In this section we list the Yukawa textures for all possible
GCP implementations in the scalar and down-type quark Yukawa sectors of the 3HDM.
(As mentioned, for the up-type quark Yukawa matrices one simply makes the replacement
$\beta \rightarrow \gamma$).
This has been performed in two independent ways.
The first used the method described in the previous section.
For the second,
we have developed an extensive Mathematica program that automatically
tests all possible GCP assignments.
This is done by solving Eqs.~\eqref{GCP_Gamma_12} and ~\eqref{GCP_Gamma_3}
through the method described in Appendix~\ref{app:scalar_matrices}.
Through this strategy, we can obtain, for each aforementioned block,
a homogeneous system of equations
$\textbf{M}(\theta, \alpha, \beta)\, \Gamma^{'}=\textbf{0}$,
where $\textbf{M}$ is a matrix which depends on the angles of the
GCP symmetry assignment and $\Gamma^{'}$ is a vector containing the
real (imaginary) part of the entries of $\Gamma_{\bar{k}}$ and
$\Gamma_3$ for each block.
To find the solutions for the system, the program evaluates
which sets of $(\theta, \alpha, \beta)$ nullify the
determinant of $\textbf{M}$ and subsequently calculates $\Gamma^{'}$
such that it is in the null space of $\textbf{M}$.
It then performs all possible logical combinations of these solutions,
as well as the trivial solution, and simplifies the result
excluding redundant and impossible cases.
Finally, it computes de eigenvalues of $H_d$ and $H_u$ as well as the
Jarlskog invariant for each possible case,
and excludes nonphysical results.

We have confirmed that Table \ref{tab:final_cases} is correct and complete.
Moreover, the program also identifies automatically all relevant Yukawa
matrices; they are shown explicitly in this section.

\subsection{CPa: consistent Yukawa textures}

CPa (the usual type of CP) implies simply that all parameters are real:
\be
\Gamma_1 =
\left(
\begin{array}{ccc}
 a_{11} & a_{12} & a_{13} \\
 a_{21} & a_{22} & a_{23} \\
 a_{31} & a_{32} & a_{33}
\end{array}
\right),
\ \ \
\Gamma_2 =
\left(
\begin{array}{ccc}
 c_{11} & c_{12} & c_{13} \\
 c_{21} & c_{22} & c_{23} \\
 c_{31} & c_{32} & c_{33}
\end{array}
\right),
\ \ \
\Gamma_3 = \left(
\begin{array}{ccc}
 e_{11} & e_{12} & e_{13} \\
 e_{21} & e_{22} & e_{23} \\
 e_{31} & e_{32} & e_{33}
\end{array}
\right).
\ee
Here and in the following subsections,
the parameters
$a_{ij}$, $b_{ij}$, $c_{ij}$, $d_{ij}$, $e_{ij}$,
and $f_{ij}$
are all real.
Imaginary numbers will be shown explicitly.

\subsection{CPb: consistent Yukawa textures}

\subsubsection{$(\theta, \alpha, \beta) = (\pi/2, 0, 0)$}

Here,
\be
\Gamma_1 = \Gamma_2 = 0,
\ \ \
\Gamma_3 =
\left(
\begin{array}{ccc}
 e_{11} & e_{12} & e_{13} \\
 e_{21} & e_{22} & e_{23} \\
 e_{31} & e_{32} & e_{33}
\end{array}
\right).
\ee

\subsubsection{$(\theta, \alpha, \beta) = (\pi/2, 0, \pi/2)$}
\ba
\Gamma_1 &=&
\left(
\begin{array}{ccc}
 a_{11}+i b_{11} & a_{12}+i b_{12} & 0 \\
 a_{21}+i b_{21} & a_{22}+i b_{22} & 0 \\
 a_{31}+i b_{31} & a_{32}+i b_{32} & 0
\end{array}
\right),
\nonumber\\
\Gamma_2 &=&
\left(
\begin{array}{ccc}
 -a_{12} + i b_{12} & a_{11}-i b_{11} & 0 \\
 -a_{22} + i b_{22} & a_{21}-i b_{21} & 0 \\
 -a_{32} + i b_{32} & a_{31}-i b_{31} & 0
\end{array}
\right),
\ \ \
\Gamma_3 =
\left(
\begin{array}{ccc}
 0 & 0 & e_{13} \\
 0 & 0 & e_{23} \\
 0 & 0 & e_{33}
\end{array}
\right).
\ea

\subsubsection{$(\theta, \alpha, \beta) = ( \pi/2, \pi/2, 0)$}
\ba
\Gamma_1 &=&
\left(
\begin{array}{ccc}
 a_{11}+i b_{11} & a_{12}+i b_{12} & a_{13}+i b_{13} \\
 a_{21}+i b_{21} & a_{22}+i b_{22} & a_{23}+i b_{23} \\
 0 & 0 & 0
\end{array}
\right),
\nonumber\\
\Gamma_2 &=&
\left(
\begin{array}{ccc}
 -a_{21} + i b_{21} & -a_{22} + i b_{22} & -a_{23} + i b_{23} \\
 a_{11}-i b_{11} & a_{12}-i b_{12} & a_{13}-i b_{13} \\
 0 & 0 & 0
\end{array}
\right),
\ \ \
\Gamma_3 =
\left(
\begin{array}{ccc}
 0 & 0 & 0 \\
 0 & 0 & 0 \\
 e_{31} & e_{32} & e_{33}
\end{array}
\right).
\ea

\subsubsection{$(\theta, \alpha, \beta) = ( \pi/2, \pi/2, \pi/2)$}
\ba
\Gamma_1 &=&
\left(
\begin{array}{ccc}
 0 & 0 & a_{13}+i b_{13} \\
 0 & 0 & a_{23}+i b_{23} \\
 a_{31}+i b_{31} & a_{32}+i b_{32} & 0
\end{array}
\right),
\nonumber\\
\Gamma_2 &=&
\left(
\begin{array}{ccc}
 0 & 0 & -a_{23} + i b_{23} \\
 0 & 0 & a_{13}-i b_{13} \\
 -a_{32} + i b_{32} & a_{31}-i b_{31} & 0
\end{array}
\right),
\ \ \
\Gamma_3 =
\left(
\begin{array}{ccc}
 e_{11}+i f_{11} & e_{12} + i f_{12} & 0 \\
 -e_{12}+i f_{12} & e_{11}-i f_{11} & 0 \\
 0 & 0 & e_{33}
\end{array}
\right).
\ea

\subsection{CPc: consistent Yukawa textures}

\subsubsection{$(\theta, \alpha, \beta) = ( \pi/3, 0, 0)$}
\be
\Gamma_1 = \Gamma_2 = 0,
\ \ \
\Gamma_3 = \left(
\begin{array}{ccc}
 e_{11} & e_{12} & e_{13} \\
 e_{21} & e_{22} & e_{23} \\
 e_{31} & e_{32} & e_{33}
\end{array}
\right).
\ee

\subsubsection{$(\theta, \alpha, \beta) = ( \pi/3, 0, \pi/3)$}
\be
\Gamma_1 = 
\left(
\begin{array}{ccc}
 a_{11} & a_{12} & 0 \\
 a_{21} & a_{22} & 0 \\
 a_{31} & a_{32} & 0
\end{array}
\right),
\ \ \
\Gamma_2 =
\left(
\begin{array}{ccc}
 -a_{12} & a_{11} & 0 \\
 -a_{22} & a_{21} & 0 \\
 -a_{32} & a_{31} & 0
\end{array}
\right),
\ \ \
\Gamma_3 =
\left(
\begin{array}{ccc}
 0 & 0 & e_{13} \\
 0 & 0 & e_{23} \\
 0 & 0 & e_{33}
\end{array}
\right).
\ee

\subsubsection{$(\theta, \alpha, \beta) = ( \pi/3, \pi/3, 0)$}
\be
\Gamma_1 =
\left(
\begin{array}{ccc}
 a_{11} & a_{12} & a_{13} \\
 a_{21} & a_{22} & a_{23} \\
 0 & 0 & 0
\end{array}
\right),
\ \ \
\Gamma_2 =
\left(
\begin{array}{ccc}
 -a_{21} & -a_{22} & -a_{23} \\
 a_{11} & a_{12} & a_{13} \\
 0 & 0 & 0
\end{array}
\right),
\ \ \
\Gamma_3 =
\left(
\begin{array}{ccc}
 0 & 0 & 0 \\
 0 & 0 & 0 \\
 e_{31} & e_{32} & e_{33}
\end{array}
\right).
\ee

\subsubsection{$(\theta, \alpha, \beta) = ( \pi/3, \pi/3, \pi/3)$}
\be
\Gamma_1 =
\left(
\begin{array}{ccc}
 i b_{11} & i b_{12} & a_{13} \\
 i b_{12} & -i b_{11} & a_{23} \\
 a_{31} & a_{32} & 0
\end{array}
\right),
\ \ \
\Gamma_2 =
\left(
\begin{array}{ccc}
 i b_{12} & -i b_{11} & -a_{23} \\
 -i b_{11} & -i b_{12} & a_{13} \\
 -a_{32} & a_{31} & 0
\end{array}
\right),
\ \ \
\Gamma_3 =
\left(
\begin{array}{ccc}
e_{11} & e_{12} & 0 \\
-e_{12} & e_{11} & 0 \\
0 & 0 & e_{33}
\end{array}
\right).
\ee

\subsection{CPd: consistent Yukawa textures}

\subsubsection{$(\theta, \alpha, \beta) = ( \pi/4, \pi/4, \pi/2)$}
\ba
\Gamma_1 &=&
\left(
\begin{array}{ccc}
a_{11} + i b_{11} & a_{12} + i b_{12} & a_{13} \\
-a_{12} + i b_{12} & a_{11} - i b_{11} & a_{23} \\
0 & 0 & 0
\end{array}
\right),
\nonumber\\
\Gamma_2 &=&
\left(
\begin{array}{ccc}
-a_{12} + i b_{12} & a_{11} - i b_{11} & -a_{23} \\
-a_{11} - i b_{11} & -a_{12} - i b_{12} & a_{13} \\
0 & 0 & 0
\end{array}
\right),
\ \ \
\Gamma_3 =
\left(
\begin{array}{ccc}
0 & 0 & 0 \\
0 & 0 & 0 \\
0 & 0 & e_{33}
\end{array}
\right).
\ea

\subsubsection{$(\theta, \alpha, \beta) = ( \pi/4, \pi/2, \pi/4)$}
\ba
\Gamma_1 &=&
\left(
\begin{array}{ccc}
a_{11} + i b_{11} & a_{12} + i b_{12} & 0 \\
-a_{12} + i b_{12} & a_{11} - i b_{11} & 0 \\
a_{31} & a_{32} & 0
\end{array}
\right),
\nonumber\\
\Gamma_2 &=&
\left(
\begin{array}{ccc}
a_{12} +i b_{12} & -a_{11} - i b_{11} & 0 \\
a_{11} - i b_{11} & a_{12} - i b_{12} & 0 \\
-a_{32} & a_{31} & 0
\end{array}
\right),
\ \ \
\Gamma_3 =
\left(
\begin{array}{ccc}
0 & 0 & 0 \\
0 & 0 & 0 \\
0 & 0 & e_{33}
\end{array}
\right).
\ea

\subsubsection{$(\theta, \alpha, \beta) = ( \pi/5, 2\pi/5, 2\pi/5)$}
\be
\Gamma_1 =
\left(
\begin{array}{ccc}
i b_{11} & i b_{12} & 0 \\
i b_{12} & -i b_{11} & 0 \\
0 & 0 & 0
\end{array}
\right),
\ \ \
\Gamma_2 =
\left(
\begin{array}{ccc}
i b_{12} & -i b_{11} & 0 \\
-i b_{11} & -i b_{12} & 0 \\
0 & 0 & 0
\end{array}
\right),
\ \ \
\Gamma_3 =
\left(
\begin{array}{ccc}
e_{11} & e_{12} & 0 \\
-e_{12} & e_{11} & 0 \\
0 & 0 & e_{33}
\end{array}
\right).
\ee

\subsubsection{$(\theta, \alpha, \beta) = ( 2\pi/5, \pi/5, \pi/5)$}
\be
\Gamma_1 =
\left(
\begin{array}{ccc}
a_{11} & a_{12} & 0 \\
a_{12} & -a_{11} & 0 \\
0 & 0 & 0
\end{array}
\right),
\ \ \
\Gamma_2 =
\left(
\begin{array}{ccc}
-a_{12} & a_{11} & 0 \\
a_{11} & a_{12} & 0 \\
0 & 0 & 0
\end{array}
\right),
\ \ \
\Gamma_3 =
\left(
\begin{array}{ccc}
e_{11} & e_{12} & 0 \\
-e_{12} & e_{11} & 0 \\
0 & 0 & e_{33}
\end{array}
\right).
\ee

\subsubsection{$\theta \notin \{ 0, \pi/3, \pi/2\}$ and $(\alpha, \beta) = ( 0, 0)$}
\be
\Gamma_1 = \Gamma_2 = 0,
\ \ \
\Gamma_3 =
\left(
\begin{array}{ccc}
e_{11} & e_{12} & e_{13} \\
e_{21} & e_{22} & e_{23} \\
e_{31} & e_{32} & e_{33}
\end{array}
\right).
\ee

\subsubsection{$\theta \notin \{ 0, \pi/3, \pi/2\}$ and $(\alpha, \beta) = ( 0, \theta)$}
\be
\Gamma_1 =
\left(
\begin{array}{ccc}
a_{11} & a_{12} & 0 \\
a_{21} & a_{22} & 0 \\
a_{31} & a_{32} & 0
\end{array}
\right),
\ \ \
\Gamma_2 =
\left(
\begin{array}{ccc}
-a_{12} & a_{11} & 0 \\
-a_{22} & a_{21} & 0 \\
-a_{32} & a_{31} & 0
\end{array}
\right),
\ \ \
\Gamma_3 =
\left(
\begin{array}{ccc}
0 & 0 & e_{13} \\
0 & 0 & e_{23} \\
0 & 0 & e_{33}
\end{array}
\right).
\ee

\subsubsection{$\theta \notin \{ 0, \pi/3, \pi/2\}$ and $(\alpha, \beta) = ( \theta, 0)$}
\be
\Gamma_1 =
\left(
\begin{array}{ccc}
a_{11} & a_{12} & a_{13} \\
a_{21} & a_{22} & a_{23} \\
0 & 0 & 0
\end{array}
\right),
\ \ \
\Gamma_2 =
\left(
\begin{array}{ccc}
-a_{21} & -a_{22} & -a_{23} \\
a_{11} & a_{12} & a_{13} \\
0 & 0 & 0
\end{array}
\right),
\ \ \
\Gamma_3 =
\left(
\begin{array}{ccc}
0 & 0 & 0 \\
0 & 0 & 0 \\
e_{31} & e_{32} & e_{33}
\end{array}
\right).
\ee

\subsubsection{$\theta \in \{ 0, \pi/3, \pi/2\}$ and $(\alpha, \beta) = ( \theta, \theta)$}
\be
\Gamma_1 =
\left(
\begin{array}{ccc}
0 & 0 & a_{13} \\
0 & 0 & a_{23} \\
a_{31} & a_{32} & 0
\end{array}
\right),
\ \ \
\Gamma_2 =
\left(
\begin{array}{ccc}
0 & 0 & -a_{23} \\
0 & 0 & a_{13} \\
-a_{32} & a_{31} & 0
\end{array}
\right),
\ \ \
\Gamma_3 =
\left(
\begin{array}{ccc}
e_{11} & e_{12} & 0 \\
-e_{12} & e_{11} & 0 \\
0 & 0 & e_{33}
\end{array}
\right).
\ee

\subsubsection{$\theta \in ( 0, \pi/4 )$ and $(\alpha, \beta) = ( \theta, 2 \theta)$}
\be
\Gamma_1 =
\left(
\begin{array}{ccc}
a_{11} & a_{12} & a_{13} \\
-a_{12} & a_{11} & a_{23} \\
0 & 0 & 0
\end{array}
\right),
\ \ \
\Gamma_2 =
\left(
\begin{array}{ccc}
-a_{12} & a_{11} & -a_{23} \\
-a_{11} & -a_{12} & a_{13} \\
0 & 0 & 0
\end{array}
\right),
\ \ \
\Gamma_3 =
\left(
\begin{array}{ccc}
0 & 0 & 0 \\
0 & 0 & 0 \\
0 & 0 & e_{33}
\end{array}
\right).
\ee

\subsubsection{$\theta \in ( 0, \pi/4 )$ and $(\alpha, \beta) = ( 2 \theta, \theta)$}
\be
\Gamma_1 =
\left(
\begin{array}{ccc}
a_{11} & a_{12} & 0 \\
-a_{12} & a_{11} & 0 \\
a_{31} & a_{32} & 0
\end{array}
\right),
\ \ \
\Gamma_2 =
\left(
\begin{array}{ccc}
a_{12} & -a_{11} & 0 \\
a_{11} & a_{12} & 0 \\
-a_{32} & a_{31} & 0
\end{array}
\right),
\ \ \
\Gamma_3 =
\left(
\begin{array}{ccc}
0 & 0 & 0 \\
0 & 0 & 0 \\
0 & 0 & e_{33}
\end{array}
\right).
\label{special_onlyone}
\ee

\subsubsection{$\theta \in ( \pi/4, \pi/2 )\setminus\{ \pi/3 \}$ and $(\alpha, \beta) = ( \theta, \pi - 2 \theta)$}
\be
\Gamma_1 =
\left(
\begin{array}{ccc}
i b_{11} & i b_{12} & a_{13} \\
i b_{12} & -i b_{11} & a_{23} \\
0 & 0 & 0
\end{array}
\right),
\ \ \
\Gamma_2 =
\left(
\begin{array}{ccc}
i b_{12} & -i b_{11} & -a_{23} \\
-i b_{11} & -i b_{12} & a_{13} \\
0 & 0 & 0
\end{array}
\right),
\ \ \
\Gamma_3 =
\left(
\begin{array}{ccc}
0 & 0 & 0 \\
0 & 0 & 0 \\
0 & 0 & e_{33}
\end{array}
\right).
\ee

\subsubsection{$\theta \in ( \pi/4, \pi/2 )\setminus\{ \pi/3 \}$
and $(\alpha, \beta) = (\pi - 2 \theta , \theta )$}
\be
\Gamma_1 =
\left(
\begin{array}{ccc}
i b_{11} & i b_{12} & 0 \\
i b_{12} & -i b_{11} & 0 \\
a_{31} & a_{32} & 0
\end{array}
\right),
\ \ \
\Gamma_2 =
\left(
\begin{array}{ccc}
i b_{12} & -i b_{11} & 0 \\
-i b_{11} & -i b_{12} & 0 \\
-a_{32} & a_{31} & 0
\end{array}
\right),
\ \ \
\Gamma_3 =
\left(
\begin{array}{ccc}
0 & 0 & 0 \\
0 & 0 & 0 \\
0 & 0 & e_{33}
\end{array}
\right).
\ee

\section{\label{sec:conclusions}Conclusions}

We have studied the implementation of a generalized CP symmetry on
the scalar and Yukawa sectors of the 3HDM. By introducing a key
mathematical result, we simplify the analysis, especially in the
Yukawa sector.

In the scalar sector, we identified four classes of potentials,
just one more than the two Higgs doublet models (2HDM).
We were also able to identify that these four potentials are
classified as CP2, CP4, $S_3 \times GCP_{\theta=\pi}$,
and $O(2) \times  CP$
in a notation close to \cite{deMedeirosVarzielas:2019rrp}.

Additionally, we categorize all possible Yukawa textures,
excluding those that result in null or degenerate quark masses,
or a null Jarlskog invariant. We found that there are 40 different
possible implementations of GCP symmetry in the Yukawa sector.
This is in contrast with the 2HDM case with generalized CP symmetries
extended to the Yukawa sector, where only two scenarios exist.
While many of these cases have a large number of parameters in the
Yukawa sector,
there are 8 models which entail only 10 parameters and,
therefore, may be more easily tested against experimental constraints.

\begin{acknowledgments}
We are grateful to I.P. Ivanov for important discussions and to
J.C. Rom\~{a}o for help with running the project.
J.P.S. acknowledges discussions with L. Lavoura and P.M. Ferreira
in very early stages.
This work is supported in part by FCT (Funda\c{a}\~{a}o para a Ci\^{e}ncia e
Tecnologia) under Contracts
CERN/FIS-PAR/0002/2021,
CERN/FIS-PAR/0008/2019,
UIDB/00777/2020,
and UIDP/00777/2020;
these projects are partially funded through POCTI (FEDER), COMPETE,
QREN, and the EU.
\end{acknowledgments}

\vspace{5ex}


\appendix
\section{\label{app:scalar_matrices}GCP constraints on the scalar potential}

In this appendix we discuss how one would show that there are four classes of
GCP-symmetric scalar potentials in the 3HDM,
using the classical method already used in \cite{Ferreira:2010bm}.
We show the result in the special basis in which $X$
has the form in Eq.~(\ref{eq:basis}).
This is to be contrasted with the much simpler and elegant method
we establish in Appendix~\ref{app:torus} and use in Section~\ref{sec:new}.

Introducing \cite{Ferreira:2009wh}
\begin{eqnarray}
\Delta Y_{ab}
&=&
Y_{ab} -
X_{\alpha a} Y_{\alpha \beta}^\ast X_{\beta b}^\ast
= \left[Y - ( X^\dagger\, Y\, X )^\ast \right]_{ab},
\nonumber\\
\Delta Z_{ab,cd}
&=&
Z_{ab,cd} -
X_{\alpha a} X_{\gamma c}
Z_{\alpha \beta, \gamma \delta}^\ast X_{\beta b}^\ast X_{\delta d}^\ast\, ,
\label{DY-DZ}
\end{eqnarray}
we may write the conditions for invariance under GCP as
\begin{eqnarray}
\Delta Y_{ab}
&=&
0,
\label{DY-GCP}
\\
\Delta Z_{ab,cd}
&=&
0.
\label{DZ-GCP}
\end{eqnarray}
Given Eqs.~(\ref{Hermiticity_coefficients}),
it is easy to show that
\begin{eqnarray}
\Delta Y_{ab}
&=&
\Delta Y_{ba}^\ast,
\nonumber\\
\Delta Z_{ab,cd} \equiv \Delta Z_{cd,ab}
&=&
\Delta Z_{ba,dc}^\ast.
\label{DY-DZ-Hermiticity}
\end{eqnarray}
%

The task of solving Eqs.~\eqref{eq:sclinv} or, what is the same,
Eqs.~\eqref{DY-GCP}-\eqref{DZ-GCP} becomes tedious but systematic.
Let us look at the quadratic terms.
We find
\ba
0
&=&
\textrm{Im} \left( \Delta Y_{12} \right) = 2 \textrm{Im} \left( Y_{12} \right),
\nonumber\\*[3mm]
\left[
\begin{array}{c}
0\\
0
\end{array}
\right]
&=&
\left[
\begin{array}{c}
\Delta Y_{11} = - \Delta Y_{22}\\
\textrm{Re} \left( \Delta Y_{12} \right)
\end{array}
\right]
=
s_\theta
\left[
\begin{array}{cc}
 s_\theta & c_\theta\\
- c_\theta & s_\theta
\end{array}
\right]
\ 
\left[
\begin{array}{c}
Y_{11} - Y_{22}\\
2 \textrm{Re} \left( Y_{12} \right)
\end{array}
\right],
\nonumber\\*[3mm]
\left[
\begin{array}{c}
0\\
0
\end{array}
\right]
&=&
\left[
\begin{array}{c}
\textrm{Re} \left( \Delta Y_{13} \right)\\
\textrm{Re} \left( \Delta Y_{23} \right)
\end{array}
\right]
=
\left[
\begin{array}{cc}
 1 - c_\theta & s_\theta\\
- s_\theta & 1 - c_\theta
\end{array}
\right]
\ 
\left[
\begin{array}{c}
\textrm{Re} \left( Y_{13} \right)\\
\textrm{Re} \left( Y_{23} \right)
\end{array}
\right],
\nonumber\\*[3mm]
\left[
\begin{array}{c}
0\\
0
\end{array}
\right]
&=&
\left[
\begin{array}{c}
\textrm{Im} \left( \Delta Y_{13} \right)\\
\textrm{Im} \left( \Delta Y_{23} \right)
\end{array}
\right]
=
\left[
\begin{array}{cc}
 1 + c_\theta & - s_\theta\\
s_\theta & 1 + c_\theta
\end{array}
\right]
\ 
\left[
\begin{array}{c}
\textrm{Im} \left( Y_{13} \right)\\
\textrm{Im} \left( Y_{23} \right)
\end{array}
\right].\label{eq:Yold}
\ea
The determinants of the three systems are 
$s_\theta^2$,
$2(1 - c_\theta)$,
and $2(1 + c_\theta)$,
respectively.
Given the restricted range $0 \leq \theta \leq \pi/2$,
the first two are zero if and only if $\theta=0$,
while the third never vanishes.
As a result,
we conclude that there are two possibilities:
either $\theta=0$
(corresponding to the usual definition
of CP, denoted here by CPa),
and all quadratic coefficients are real;
or $\theta \neq 0$,
and $Y = \textrm{diag} (\mu_1, \mu_1, \mu_3)$,
with $\mu_1$ and $\mu_3$ real.
The first case corresponds to
6 real parameters;
the second case corresponds to
2 real parameters.

The study of the GCP symmetry conditions on the quartic couplings using
this strategy is more involved.
To see the impact on the quartic potential,
we follow Ref.~\cite{Ferreira:2008zy} and
organize the $Z_{ij,kl}$ tensor into
a matrix of matrices.
The uppermost-leftmost matrix corresponds to
the phases affecting $Z_{11,kl}$. 
The next matrix along the same line corresponds
to the phases affecting $Z_{12,kl}$, and so on...
We use the following notation for the various
entries \cite{Ferreira:2008zy}
\begin{equation}
\left[
\begin{array}{ccc}
    \left[
    \begin{array}{ccc}
      r_1 & c_1 & c_2 \\
      c_1^\ast & r_4 & c_6 \\
      c_2^\ast & c_6^\ast & r_5 \\
    \end{array}
    \right]
&
    \left[
    \begin{array}{ccc}
      c_1 & c_3 & c_4 \\
      r_7 & c_7 & c_8 \\
      c_9^\ast & c_{12} & c_{13} \\
    \end{array}
    \right]
&
    \left[
    \begin{array}{ccc}
      c_2 & c_4 & c_5 \\
      c_9 & c_{10} & c_{11} \\
      r_8 & c_{14} & c_{15} \\
    \end{array}
    \right]
\\*[12mm]
    \left[
    \begin{array}{ccc}
      c_1^\ast & r_7 & c_9 \\
      c_3^\ast & c_7^\ast & c_{12}^\ast \\
      c_4^\ast & c_8^\ast & c_{13}^\ast \\
    \end{array}
    \right]
&
    \left[
    \begin{array}{ccc}
      r_4 & c_7 & c_{10} \\
      c_7^\ast & r_2 & c_{16} \\
      c_{10}^\ast & c_{16}^\ast & r_6 \\
    \end{array}
    \right]
&
    \left[
    \begin{array}{ccc}
      c_6 & c_8 & c_{11} \\
      c_{12}^\ast & c_{16} & c_{17} \\
      c_{14}^\ast & r_9 & c_{18} \\
    \end{array}
    \right]
\\*[12mm]
    \left[
    \begin{array}{ccc}
      c_2^\ast & c_9^\ast & r_8 \\
      c_4^\ast & c_{10}^\ast & c_{14}^\ast \\
      c_5^\ast & c_{11}^\ast & c_{15}^\ast \\
    \end{array}
    \right]
&
    \left[
    \begin{array}{ccc}
      c_6^\ast & c_{12} & c_{14} \\
      c_8^\ast & c_{16}^\ast & r_9 \\
      c_{11}^\ast & c_{17}^\ast & c_{18}^\ast \\
    \end{array}
    \right]
&
    \left[
    \begin{array}{ccc}
      r_5 & c_{13} & c_{15} \\
      c_{13}^\ast & r_6 & c_{18} \\
      c_{15}^\ast & c_{18}^\ast & r_3 \\
     \end{array}
    \right]
\\
\end{array}
\right],
\label{quartic}
\end{equation}
where $r_k$ ($k = 1 \dots 9$) are real and $c_k$ ($k = 1 \dots 18$) 
are complex.
We will write $c_k = x_k + i y_k$,
with $x_k$ and $y_k$ real.

We now wish to study Eqs.~\eqref{DZ-GCP}.
Due to Eqs.~(\ref{DY-DZ-Hermiticity}),
we only need the 9 real coefficients
$\Delta Z_{11,11}$, $\Delta Z_{22,22}$, $\Delta Z_{33,33}$,
$\Delta Z_{11,22}$, $\Delta Z_{11,33}$, $\Delta Z_{22,33}$,
$\Delta Z_{12,21}$, $\Delta Z_{13,31}$, $\Delta Z_{23,32}$,
and the 18 complex coefficients
$\Delta Z_{11,12}$, $\Delta Z_{11,13}$, $\Delta Z_{11,23}$,
$\Delta Z_{22,12}$, $\Delta Z_{22,13}$, $\Delta Z_{22,23}$,
$\Delta Z_{33,12}$, $\Delta Z_{33,13}$, $\Delta Z_{33,23}$,
$\Delta Z_{12,12}$, $\Delta Z_{12,13}$, $\Delta Z_{12,23}$, 
$\Delta Z_{12,31}$, $\Delta Z_{12,32}$,
$\Delta Z_{13,13}$, $\Delta Z_{13,23}$,
$\Delta Z_{13,32}$,
and $\Delta Z_{23,23}$.
However,
\ba
- 2 \Delta Z_{11,22}
&=&
\Delta Z_{11,11} + \Delta Z_{22,22}
\nonumber\\
\Delta Z_{22,33}
&=&
- \Delta Z_{11,33},
\nonumber\\
\Delta Z_{33,33}
&=&
0,
\nonumber\\
\Delta Z_{12,21}
&=&
\Delta Z_{11,22},
\nonumber\\
\Delta Z_{23,32}
&=&
- \Delta Z_{13,31},
\nonumber\\
\textrm{Re} \left( \Delta Z_{12,12} \right)
&=&
\textrm{Re} \left( \Delta Z_{11,22} \right),
\nonumber\\
\textrm{Re} \left( \Delta Z_{23,23} \right)
&=&
- \textrm{Re} \left( \Delta Z_{13,13} \right),
\ea
simplifying the analysis.
One then proceeds as for the quadratic terms finding systems such as, for example,
\ba
\left[
\begin{array}{c}
0\\
0
\end{array}
\right]
&=&
\left[
\begin{array}{c}
\textrm{Im} \left( \Delta Z_{12,13} - \Delta Z_{11,23} \right) \\
\textrm{Im} \left( \Delta Z_{22,13} + \Delta Z_{12,32} \right)
\end{array}
\right]
=
\left[
\begin{array}{cc}
 1 + c_\theta & - s_\theta\\
s_\theta & 1 + c_\theta
\end{array}
\right]
\ 
\left[
\begin{array}{c}
y_4 - y_6 \\
y_{10} + y_{12}
\end{array}
\right],
\nonumber\\*[3mm]
\left[
\begin{array}{c}
0\\
0
\end{array}
\right]
&=&
\left[
\begin{array}{c}
\textrm{Im} \left( \Delta Z_{11,23} + \Delta Z_{12,31} \right) \\
\textrm{Im} \left( \Delta Z_{12,23} - \Delta Z_{22,13} \right)
\end{array}
\right]
=
\left[
\begin{array}{cc}
 1 + c_\theta & - s_\theta\\
s_\theta & 1 + c_\theta
\end{array}
\right]
\ 
\left[
\begin{array}{c}
y_6 - y_9 \\
y_{8} - y_{10}
\end{array}
\right],
\nonumber\\*[3mm]
\left[
\begin{array}{c}
0\\
0
\end{array}
\right]
&=&
\left[
\begin{array}{c}
\textrm{Im} \left( \Delta Z_{33,13} \right) \\
\textrm{Im} \left( \Delta Z_{33,23} \right)
\end{array}
\right]
=
\left[
\begin{array}{cc}
 1 + c_\theta & - s_\theta\\
s_\theta & 1 + c_\theta
\end{array}
\right]
\ 
\left[
\begin{array}{c}
y_{15} \\
y_{18}
\end{array}
\right]
\ea
These systems have determinant $2(1+c_\theta)$,
which never vanishes due to the restricted range $0 \leq \theta \leq \pi/2$.
As a result,
any GCP symmetry will force
$y_{15}=y_{18}=0$,
$y_4=y_6=y_9$,
and
$y_{8}=y_{10}=- y_{12}$.

A more challenging example is
\ba
\left[
\begin{array}{c}
0\\
0\\
0\\
0
\end{array}
\right]
&=&
\left[
\begin{array}{c}
\textrm{Im} \left( \Delta Z_{11,13} \right) \\
4 \textrm{Im} \left( \Delta Z_{12,13} + \Delta Z_{11,23} - \Delta Z_{12,31} \right)\\
4 \textrm{Im} \left( \Delta Z_{12,23} + \Delta Z_{22,13} - \Delta Z_{12,32} \right)\\
\textrm{Im} \left( \Delta Z_{22,23} \right)
\end{array}
\right]
\nonumber\\
&=&
\left[
\begin{array}{cccc}
 1+c_{\theta }^3 & -c_{\theta }^2 s_{\theta } & c_{\theta } s_{\theta }^2 & -s_{\theta }^3 \\
 12 c_{\theta }^2 s_{\theta } & 4+ c_{\theta }+3 c_{3 \theta } & s_{\theta }-3 s_{3 \theta } & 12 c_{\theta }
   s_{\theta }^2 \\
 12 c_{\theta } s_{\theta }^2 & 3 s_{3 \theta }-s_{\theta } & 4+c_{\theta }+3 c_{3 \theta } & -12 c_{\theta }^2
   s_{\theta } \\
 s_{\theta }^3 & c_{\theta } s_{\theta }^2 & c_{\theta }^2 s_{\theta } & 1+c_{\theta }^3
\end{array}
\right]
\ 
\left[
\begin{array}{c}
y_2 \\
y_4 + y_6 + y_9\\
y_8 + y_{10} - y_{12}\\
y_{16}
\end{array}
\right],
\ea
with determinant $64  (1+c_{\theta })^2 (1-2 c_{\theta })^2$.
Given the restricted range $0 \leq \theta \leq \pi/2$,
this determinant can only vanish if $c_\theta = 1/2$,
that is, $\theta=\pi/3$.

We have checked explicitly that this procedure does reproduce the results
for the scalar potential obtained in the main text, although
in a much more tedious way.

\section{\label{app:torus}A mathematical result on Kroenecker products of rotations}

In this appendix we state and use
a result that greatly simplifies systems of equations involving tensors and rotation matrices. 
If we work in the scalar field basis where the GCP transformation matrix is of the form
in Eq.~\eqref{eq:basis}, with $\oplus$ denoting the direct sum operation,
we can use this result to solve the equations regarding GCP-invariance
in both the scalar and Yukawa sectors of the SM Lagrangian with great ease.

\textbf{ \underline{Result}:}
Suppose we have a $2^n\times2^n$ matrix, with $n\in\mathbb{N}$, defined as:
\begin{equation}
    \mathbf{R}_\Theta^n \equiv \bigotimes_{q=1}^{n}R_{\theta_q}
= R_{\theta_1}\otimes R_{\theta_2}\otimes\cdots\otimes R_{\theta_n}\, .
\label{result}
\end{equation}
\begin{equation}
\begin{split}
&
\text{Then: } \forall n\geq 1 \;\exists C_n \in M_{2^n\times2^n}:\ 
 C_n C_n^\top=I_{2^n}\, ,\ \textrm{and}\\
&
\hspace{20ex}
C_n \mathbf{R}_\Theta^n C_n^\top 
= \bigoplus_{p=0}^{2^{n-1}-1} R_{\omega_p^n}
= R_{\omega_0^n} \oplus R_{\omega_1^n} \oplus\cdots\oplus
R_{\omega_{2^{n-1}-1}^n}\; \text{, }
\end{split}
\label{BB2}
\end{equation}
where
%
\begin{equation}
\omega_p^n = \displaystyle\sum_{q=1}^n (-1)^{\textrm{Mod}(\lfloor \frac{p}{2^{q-1}}\rfloor,2)}\theta_q\, ,
\label{BB3}
\end{equation}
$I_{2^n}$ is the identity
$2^n \times 2^n$ matrix,
and $\lfloor \ \rfloor$ is the floor function, 
which yields the previous largest integer. 
For example, for $n=3$ we have $\omega_0^3=\theta_1+\theta_2+\theta_3$,
$\omega_1^3=-\theta_1+\theta_2+\theta_3$,
$\omega_2^3=\theta_1-\theta_2+\theta_3$,
and $\omega_3^3=-\theta_1-\theta_2+\theta_3$.

The $C_n$ matrices are defined as follows:
\begin{equation}
    C_n= \left(\bigoplus_{p=1}^{2^n}U^{-1}\right)\left( P_n \right)\left(\bigotimes_{p=1}^{n}U \right)\,, \label{eq:finaleq}
\end{equation}
where
\begin{eqnarray}
&&
(P_n)_{pq}=1\, , \quad \textrm{if }\ 
\left\{
\begin{array}{lcl}
p \textrm{ is odd} & \textrm{ and }  & q = \frac{p+1}{2} \\
p \textrm{ is even} & \textrm{ and }  & q = 2^n - \frac{p}{2} +1 \\
\end{array}
\right.
\, ,
\nonumber\\*[2mm]
&&
(P_n)_{pq}=0\, ,\ \ \ \textrm{otherwise}\, ,
\label{def:Pn}
\end{eqnarray}
and
\begin{equation}
    U=\frac{1}{\sqrt{2}}\begin{pmatrix}
        1 & i \\
        1 & -i 
    \end{pmatrix}\,.
\end{equation}

We include in the supplemental material a proof of
Eqs.~\eqref{BB2}-\eqref{eq:finaleq}.
In addition,
using
the first few $C_n$ matrices,
\begin{equation}
    C_2 = \frac{1}{\sqrt{2}}\begin{pmatrix}
        1 & 0 & 0 &-1 \\
        0 & 1 & 1 & 0 \\
        1 & 0 & 0 & 1 \\
        0 & 1 & -1& 0 
    \end{pmatrix}
\end{equation}
\begin{equation}
    C_3 = \frac{1}{{2}}\begin{pmatrix}
        1 & 0 & 0 &-1 &0 &-1&-1&0 \\
        0 & 1 & 1 & 0 &1 &0 &0 &-1\\
        1 & 0 & 0 &-1 &0 &1 &1 &0 \\
        0 & 1 & 1 & 0 &-1&0 &0 &1 \\  
        
        1 & 0 & 0 & 1 &0 &-1&1 &0 \\
        0 & 1 & -1& 0 &1 &0 &0 &1 \\
        1 & 0 & 0 & 1 &0 &1 &-1&0 \\
        0 & 1 &-1 & 0 &-1&0 &0 &-1
    \end{pmatrix}
\end{equation}
\begin{equation}
    C_4 = \frac{1}{\sqrt{8}}
\left(
\begin{array}{cccccccccccccccc}
        1 & 0 & 0 &-1 &0 &-1&-1&0   &  0 &-1 & -1& 0 &-1&0 &0 & 1\\
        0 & 1 & 1 & 0 &1 &0 &0 &-1  &  1 & 0 & 0 &-1 &0 &-1&-1& 0\\
        1 & 0 & 0 &-1 &0 &-1&-1&0   &  0 & 1 & 1 & 0 &1 &0 &0 &-1\\
        0 & 1 & 1 & 0 &1 &0 &0 &-1  &  -1& 0 & 0 & 1 &0 & 1& 1& 0\\
        
        1 & 0 & 0 &-1 &0 &1 &1 &0   &  0 &-1 &-1 & 0 & 1&0 &0 &-1\\
        0 & 1 & 1 & 0 &-1&0 &0 &1   &  1 & 0 & 0 &-1 &0 &1 &1 & 0\\
        1 & 0 & 0 &-1 &0 &1 &1 &0   &  0 & 1 & 1 & 0 &-1&0 &0 & 1\\
        0 & 1 & 1 & 0 &-1&0 &0 &1   &  -1& 0 & 0 & 1 &0 &-1&-1& 0\\  
        
        1 & 0 & 0 & 1 &0 &-1&1 &0   &  0 &-1 & 1 & 0 &-1&0 &0 &-1\\
        0 & 1 & -1& 0 &1 &0 &0 &1   &  1 & 0 & 0 & 1 &0 &-1&1 & 0\\
        1 & 0 & 0 & 1 &0 &-1&1 &0   &  0 & 1 &-1 & 0 &1 &0 &0 & 1\\
        0 & 1 & -1& 0 &1 &0 &0 &1   &  -1& 0 & 0 &-1 &0 & 1&-1& 0\\ 
        
        1 & 0 & 0 & 1 &0 &1 &-1&0   &  0 &-1 & 1 & 0 & 1&0 &0 & 1\\
        0 & 1 &-1 & 0 &-1&0 &0 &-1  &  1 & 0 & 0 & 1 &0 &1 &-1& 0\\
        1 & 0 & 0 & 1 &0 &1 &-1&0   &  0 & 1 &-1 & 0 &-1&0 &0 &-1\\
        0 & 1 &-1 & 0 &-1&0 &0 &-1  &  -1& 0 & 0 &-1 &0 &-1& 1& 0
\end{array}
\right)\, ,
\end{equation}
we have confirmed by explicit construction of the corresponding matrices
that Eqs.~\eqref{BB2}-\eqref{eq:finaleq} hold up to the highest $n=4$
case needed for our article.

\end{document}